\documentclass[sigconf]{acmart}

\usepackage{amsmath,amsfonts,bm}

\def\eqref#1{equation~\ref{#1}}

\def\1{\bm{1}}

\DeclareMathAlphabet{\mathsfit}{\encodingdefault}{\sfdefault}{m}{sl}
\SetMathAlphabet{\mathsfit}{bold}{\encodingdefault}{\sfdefault}{bx}{n}

\usepackage[utf8]{inputenc} 
\usepackage[T1]{fontenc}    
\usepackage{hyperref}       
\usepackage{url}            
\usepackage{booktabs}       
\usepackage{amsfonts}       
\usepackage{amsthm}

\newtheorem{theorem}{Theorem}[section]   

\theoremstyle{definition}

\theoremstyle{remark}

\usepackage{amsmath}        
\usepackage{nicefrac}       
\usepackage{microtype}      
\usepackage{xcolor}         
\usepackage{graphicx}
\usepackage{enumitem}
\usepackage{multirow}
\usepackage{graphicx}
\usepackage{subcaption}
\usepackage[normalem]{ulem}
\useunder{\uline}{\ul}{}
\newcommand{\ourmethod}{\texttt{\textbf{BLUE}}}
\AtBeginDocument{%
  }

\copyrightyear{2026}
\acmYear{2026}
\setcopyright{cc}
\setcctype{by}
\acmConference[WWW '26] {Proceedings of the ACM Web Conference 2026}{April 13--17, 2026}{Dubai, United Arab Emirates.}
\acmBooktitle{Proceedings of the ACM Web Conference 2026 (WWW '26), April 13--17, 2026, Dubai, United Arab Emirates}
\acmPrice{}
\acmISBN{979-8-4007-2307-0/2026/04}
\acmDOI{10.1145/3774904.3793010}
\begin{document}

\title{Genomic-Informed Heterogeneous Graph Learning for Spatiotemporal Avian Influenza Outbreak Forecasting}

\author{Jing Du}
\orcid{0000-0003-4113-0875}
\affiliation{
    \institution{Computer Science and Engineering, Faculty of Engineering}
    \institution{The University of New South Wales}
    \city{Sydney}
    \country{Australia}
}
\email{jing.du2@unsw.edu.au}

\author{Haley Stone}
\orcid{0000-0003-1997-0042}
\affiliation{
    \institution{The Kirby Institute} 
    \institution{Faculty of Medicine \& Health}
    \institution{The University of New South Wales}
    \city{Sydney}
    \country{Australia}
}
\email{h.stone@unsw.edu.au}

\author{Yang Yang}
\orcid{0000-0003-3960-5739}
\affiliation{
    \institution{Computer Science and Engineering, Faculty of Engineering}
    \institution{The University of New South Wales}
    \city{Sydney}
    \country{Australia}
}
\email{yang.yang26@unsw.edu.au}

\author{Ashna Desai}
\orcid{0009-0003-3891-2486}
\affiliation{
    \institution{Computer Science and Engineering, Faculty of Engineering}
    \institution{The University of New South Wales}
    \city{Sydney}
    \country{Australia}
}
\email{ashna.desai@student.unsw.edu.au}

\author{Hao Xue}
\orcid{0000-0003-1700-9215}
\affiliation{
    \institution{The Hong Kong University of Science and Technology (Guangzhou)}
    \city{Guangzhou}
    \country{China}
}
\email{haoxue@hkust-gz.edu.cn}

\author{Andreas Züfle}
\orcid{0000-0001-7001-4123}
\affiliation{
    \institution{Department of Computer Science}
    \institution{Emory University}
    \city{Atlanta}
    \country{United States}
}
\email{azufle@emory.edu}

\author{C. Raina MacIntyre}
\orcid{0000-0002-3060-0555}
\affiliation{
    \institution{The Kirby Institute} 
    \institution{Faculty of Medicine \& Health}
    \institution{The University of New South Wales}
    \city{Sydney}
    \country{Australia}
}
\email{r.macintyre@unsw.edu.au}

\author{Flora D. Salim}
\orcid{0000-0002-1237-1664}
\affiliation{
    \institution{Computer Science and Engineering, Faculty of Engineering}
    \institution{The University of New South Wales}
    \city{Sydney}
    \country{Australia}
}
\email{flora.salim@unsw.edu.au}

\renewcommand{\shortauthors}{Jing Du et al.}

\begin{abstract}
    Accurate forecasting of Avian Influenza Virus (AIV) outbreaks within wild bird populations necessitates models that account for complex, multi-scale transmission patterns driven by diverse factors. While conventional spatiotemporal epidemic models are robust for human-centric diseases, they rely on \textbf{spatial homophily} and \textbf{diffusive transmission} between geographic regions. This simplification is incomplete for AIV as it neglects valuable genomic information critical for capturing dynamics like high-frequency reassortment and lineage turnover at the case level (e.g., genetic descent across regions), which are essential for understanding AIV spread. To address these limitations, we systematically formulate the AIV forecasting problem and propose \ourmethod~(\textbf{B}i-\textbf{L}ayer genomic-aware heterogeneous graph f\textbf{U}sion pipelin\textbf{E}). This pipeline integrates genetic, spatial, and ecological data to achieve highly accurate outbreak forecasting. It 1) defines a multi-layered graph structure incorporating information from diverse sources and multiple layers (case and location), 2) applies cross-relation smoothing to smooth information flow across edge types, 3) performs graph fusion that preserves critical structural patterns backed by theoretical spectral guarantees, and 4) forecasts future outbreaks using an autoregressive graph sequence model to capture transmission dynamics.
    To support research, we release the Avian-US dataset, which provides comprehensive genetic, spatial, and ecological data on US avian influenza outbreaks. BLUE demonstrates superior performance over existing baselines, highlighting the efficacy of integrating multi-layer information for infectious disease forecasting.
    The code is available at: \url{https://github.com/cruiseresearchgroup/BLUE}.
\end{abstract}



\maketitle

\section{Introduction}
Forecasting the transmission of Avian Influenza Virus (AIV) is a major challenge in epidemiology due to its potential for widespread dissemination in avian populations.
The growing incidence of cross-species transmission poses substantial risks to public health and global biosecurity.
Consequently, accurate prediction of outbreak locations is vital for initiating early interventions to minimize infection risk~\citep{caliendo2022transatlantic, prosser2024using}.
Prior epidemiological models primarily utilized mechanistic approaches~\citep{geng2021kernel, della2023sir} based on biological assumptions and fixed compartmental structures.
While suitable for simplified contexts, these models fail to incorporate network topology or interaction semantics \citep{hunter2022understanding}.
Their reliance on low-dimensional Ordinary Differential Equation (ODE) systems tracks only the aggregate number of infected individuals within discrete states. This limits their ability to model temporal dependencies and the complex inter-location influence of real-world disease transmission dynamics.

To overcome the limitations of traditional epidemiological models, recent methods incorporate Graph Neural Networks (GNNs) with temporal architectures to capture spatio-temporal transmission dynamics from observational data~\citep{liu2024review}. 
These models are designed for human infectious diseases, such as Influenza-Like Illness (ILI) or COVID-19.
Their design identifies topological transmission patterns across time steps by relying on geographical correlations for connectivity.
It enables the learning of temporal correlations among spatially dispersed locations embedded in the graph structure.
This formulation offers a flexible, data-driven framework for supporting disease surveillance and control decisions, representing a significant advancement over classic forecasting methods~\citep{bruel2020universal,liu2024review}.

Nevertheless, existing spatio-temporal methods are unsuitable for AIV forecasting, as the problem formulation fundamentally differs from that of human infectious diseases.
Conventional ILI and COVID-19 forecasting address \textbf{single-host problems}.
They model infection risk as a \textit{spatio-temporal process} that propagates diffusively between adjacent locations or via human mobility networks \citep{pei2018forecasting, tang2023enhancing, kim2024forecasting}. 
In this context, a single-layer location graph is sufficient since proximity and mobility accurately proxy infection pathways among human.
By contrast, AIV spread is inherently \textbf{multi-source} and \textbf{multi-level}~\cite{venkatramanan2021forecasting, giacinti2024transmission}.
It spreads through \textit{multiple concurrent pathways}, including \textit{long-range geographical seeding} by migratory wild birds and \textit{evolutionary reassortment }\cite{kandeil2023rapid, caliendo2022transatlantic}.
Critically, these dynamics operate on \textbf{distinct spatial and genetic levels}.
For instance, two geographically distant cases can be epidemiologically identical if they share a specific viral lineage, while neighboring cases might be unrelated outbreaks.
These invisible linkages reside in the genetic space and are unobservable in the spatial domain. Models treating all cases as homogeneous count data inevitably obscure this critical genetic signal \citep{deng2020cola, tang2023enhancing}.
Consequently, this complexity reveals a mismatch between AIV dynamics and the conventional modeling paradigm.
These models typically follow a homogeneous setup, where nodes represent locations connected by static adjacencies \citep{liu2023human, yu2023spatio, lin2023graph} or learned correlations \citep{nguyen2023predicting, pu2024dynamic}. 
They focus solely on location-level transmission and treat all infected cases as equally infectious, limiting them to capturing only spatial transmission pathways.
Even recent efforts that integrate ecological variables \citep{lim2021temporal, papagiannopoulou2024long} or combine GNNs with mechanistic components \citep{cao2022mepognn, wang2022causalgnn, sha2021source} still operate on a homogeneous graph structure based on spatial locations.
This paradigm overlooks high-frequency reassortment and lineage turnover driven by case-to-case transmission, which are essential for comprehensive AIV spread understanding.

Conceptually, provides a lineage view: genetic similarity between cases suggests epidemiological linkages that transcend geography, revealing lineage-specific differences in transmissibility that are not visible through spatial or ecological proximity alone.
Existing homogeneous GNNs are inherently limited in capturing these multi-factor interactions. 
However, Heterogeneous GNNs (HGNNs) \citep{zhang2019heterogeneous} possess the capability to model heterogeneous node types (e.g., cases and locations) and construct multi-type edges based on both spatial and genetic relations, making them highly appropriate for sophisticated epidemiological modeling.
Nonetheless, existing HGNNs \citep{hemker2024healnet, kim2023heterogeneous, yu2022healthnet, guo2023graph} assumes static node sets and a fixed relational schema. 
The AIV forecasting problem violates this assumption due to two dynamics: 
1) \textbf{Hierarchical Relationship}: Cases and locations have an inherent hierarchical relationship, where each case belongs to a specific location; 
2) \textbf{Dynamic Evolution}: Each timestep involves a fixed set of location nodes but a dynamically changing set of infection cases (new infectious cases appear, dead ones vanish). This results in a dynamic evolution of both the case node set and the connectivity structure.
Existing HGNNs cannot simultaneously handle these dynamic graphs with changing node sets and distinct inter-layer semantics.
Therefore, a principled hierarchical heterogeneous graph modeling method is required to manage multi-layer heterogeneous graphs while maintaining structural integrity and semantic distinctions across layers.

To this end, we systematically ormulate AIV forecasting as a spatio-temporal hierarchical forecasting task and propose \ourmethod~($\mathbf{B}$i-$\mathbf{L}$ayer genomic-aware Graph F$\mathbf{U}$sion pipelin$\mathbf{E}$).
\ourmethod~defines cases and locations as heterogeneous nodes in dual layers and integrates spatial, genetic, and ecological information into a unified framework for outbreak prediction.
The approach begins by constructing a bi-layer heterogeneous graph with diverse nodes and multi-type edges. To manage node heterogeneity, we apply a cross-layer smoothing block, inspired by Markov Random Fields (MRF) \citep{dobruschin1968description}, to smooth the heterogeneous message passing.
Since the ultimate goal is location-based, we we then construct homogeneous fusion graphs at the location level.
We employ a Locality-Sensitive Hashing (LSH)-based sampler \citep{datar2004locality, jafari2021survey} to achieve efficient information integration from the complex bi-layer structure.
Integrating multi-type relations can lead to significant information loss in transmission structures without reasonable guarantee of preserving structural information within the graphs. 
To preserve fine-grained structural semantics, we design a spectral regularizer that constrains the learned fusion graph to approximate the global diffusion geometry of the original bi-layer structure with a theoretical bound.
To support rigorous evaluation and motivate further research, we publicly release a new avian influenza surveillance dataset.
Our main contributions are:
\begin{enumerate}[label=\arabic*., leftmargin=12pt, labelindent=0pt]
\item \textit{New Topological Paradigm}: We systematically define AIV transmission on spatial and genetic levels and propose BLUE, a pipeline that simultaneously models heterogeneous nodes with multi-type information, establishing a principled formulation beyond prior homogeneous work.
\item \textit{Theoretical guarantees for Graph Fusion}: We design an information-preserving graph fusion method to simplify the heterogeneous structure without discarding crucial epidemiological information, guaranteed by a spectral theoretical bound.
\item \textit{New AIV Dataset}: We publicly release the United States avian influenza surveillance data, Avian-US, and empirically validate BLUE, demonstrating its superior performance.
\end{enumerate}

\vspace{-2mm}
\section{Related Works}
\label{apd:related}
\paragraph{Epidemic Modeling. }
Epidemiological forecasting mainly rely on fixed graph structures, typically incorporating predefined spatial or mobility-based priors such as geographic adjacency matrices \citep{xie2022epignn, yu2023spatio, lin2023graph} or static population flow matrices \citep{liu2023human, tang2023enhancing} as connectivity structure.
For example, Cola-GNN \citep{deng2020cola} and EpiGNN \citep{xie2022epignn} combines a static region-level graph with temporal modeling via spatio-temporal graph learning.
STEP \citep{yu2023spatio} and SMPNN \citep{lin2023graph} leverage graph neural networks to perform spatio-temporal forecasting on predefined location-level graphs constructed from geographic distances. 
MSDNet \citep{tang2023enhancing} defines the graph structure based on coarse-grained population migration trajectories and employs spatio-temporal graph learning to enhance prediction. 
To capture epidemological dynamics, specifical models adopt infectious disease dynamic models (SIR/SEIR) to characterize transmission within and across regions.
Epi-Cola-GNN \citep{liu2023epidemiology} builds on this by incorporating SIS dynamics and using a learnable transmission matrix to form time-varying graphs, better reflecting real-world epidemic progression.
MepoGNN \citep{cao2022mepognn}, on the other hand, explicitly integrates SIR dynamics into the graph learning process, allowing it to model evolving infectious connections more directly. 
Despite their flexibility, these models still model disease transmission pathways only from geographical correlations, focusing on a single relation view.
%
\vspace{-2mm}
\paragraph{Spatio-temporal Forecasting. }
Spatio-temporal forecasting models often employ modules to jointly model spatial correlation and temporal dependence on irregular spatial structures.
Classic STGCN~\citep{yu2018spatio} employs spectral graph convolutional structure to capture spatial information diffusion and gated 1D temporal convolutions to learn temporal patterns.
Inspired by STGCN, SelfAttnGNN~\citep{cheng2016long} employs self-attentional networks along spatial and temporal dimensions simultaneously, enhancing long-range dependencies modeling ability.
DCRNN~\citep{li2018diffusion} integrates diffusion convolution with scheduled sampling to capture spatial and temporal evolutions.
They overlook heterogeneous factors, such as genetic relationships and evolving nodes, which are critical for understanding the multifaceted nature of real-world AIV disease transmission.
Considering multimodal interactions within graphs, EAST-Net~\citep{wang2022event} decouples spatiotemporal forecastiong into single spatial and intermodality views and proposes event-aware spatio-temporal network to jointly represent spatio-temporal dependency and intermodal interaction.

\vspace{-2mm}
\section{Methodology}
\label{sec:method}

\begin{figure*}[t]
    \centering
    \includegraphics[width=.9\linewidth]{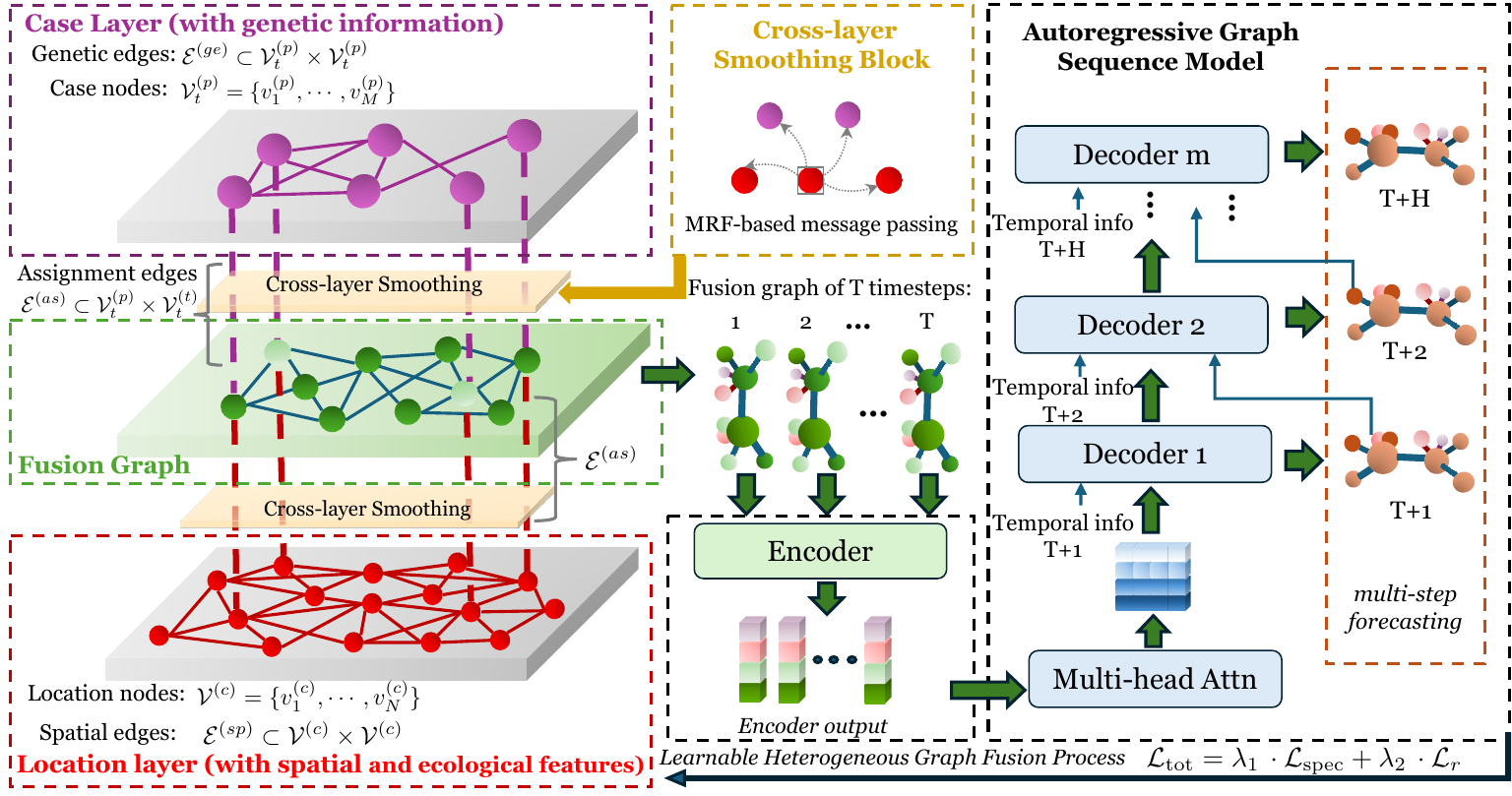}
    \vspace{-3mm}
    \caption{\ourmethod~consists of 4 components: \textit{Bi-layer Heterogeneous Graph Construction} models AIV spread using a bi-layer heterogeneous graph with two types of nodes (location and case) and three types of edges (spatial, genetic, and assignment).
    Then, the MRF-inspired \textit{Cross-layer Smoothing block}  aggregates neighbor information to create coherent representations for heterogeneous nodes and their connections.
    Graphs are then fused into \textit{Information-Preserving Fusion Graphs} that preserve the original transmission structure using a spectral regularizer.
    Finally, \textit{Autoregressive Encoder–Decoder Forecasting} encodes node interactions over time to generate multi-step forecasts.}
    \vspace{-3mm}
    \label{fig:heterograph}
\end{figure*}

\paragraph{Problem Definition.}
We formally define AIV forecasting as predicting future outbreaks using historical multi-source observations structured as a sequence of heterogeneous graphs.
Let $\mathcal{G}_t = (\mathcal{T}_V, \mathcal{T}_E)$ be the heterogeneous graph snapshot at week $t$. The node set $\mathcal{T}_V=\{ \mathcal{V}^{(c)}, \mathcal{V}^{(p)}_t\}$ comprised of two distinct types: \textbf{location nodes} $\mathcal{V}^{(c)}$ (fixed across time) and case nodes $\mathcal{V}^{(p)}_t$ (dynamic, reported infections at time $t$). 
The edge set $\mathcal{T}_E$ comprises three types of relations:
\textit{Spatial edges} (red lines) connect neighboring location nodes $\mathcal{V}^{(c)} \times \mathcal{V}^{(c)}$.
The weight $\omega_{ij}^{sp}$ is based on geographic distance.
\textit{Genetic edges} $\mathcal{E}^{(ge)} \subset \mathcal{V}_t^{(p)} \times \mathcal{V}_t^{(p)}$ connect genetically similar cases based on biological modality (shown as purple lines).
Each edge $\langle v_k^{(p)}, v_l^{(p)}\rangle$ carries a similarity weight $w_{kl}^{(ge)}$ reflecting the genetic similarity between case $k$ and case $l$.
\textit{Assignment edges} $\mathcal{E}^{(as)} \subset \mathcal{V}_t^{(p)} \times \mathcal{V}^{(c)}$ link each case to the specific location in which it was reported (shown as purple and red dashed lines).
Given the historical sequence of dynamic heterogeneous graphs over $T$ weeks, the goal is to predict the new infection counts $\mathbf{y}$ for all locations over the subsequent $H$ weeks $\{\mathbf{y}_{T+1}, \cdots, \mathbf{y}_{T+H}\} = f_{\theta}(\mathcal{G}_{1}, \cdots, \mathcal{G}_{T})$.
\ourmethod~aims to learn the function $f_{\theta}$ that accurately maps the historical graph sequence to the future location-specific case counts.

\vspace{-2mm}
\paragraph{Bi-layer Heterogeneous Graph Construction.}
Conventional spatio-temporal forecasting models rely on a single-view representation that captures only static spatial and temporal correlations between locations. 
A key weakness of these models is the assumption that all infected cases contribute equally to transmission intensity, failing to account for cases with differing transmission potential.
To address this, we propose a novel bi-layer heterogeneous graph. 
This structure simultaneously captures case-level infection intensity and location-level spatial connectivity within one unified framework.

As depicted in Fig.\ref{fig:heterograph}, the proposed bi-layer graph structure consists of two layers: \textbf{Location Layer} and \textbf{Case Layer}.

\textbf{Location Layer. }
We define a fixed set of $N$ location nodes $\mathcal{V}^{(c)}=\{ v^{(c)}_1, \cdots, v^{(c)}_N\}$ representing distinct geographic locations. 
Each location node $v^{(c)}_i$ at week $t$ is associated with a feature vector $\mathbf{x}_i^{(c)}(t) = [infected_i(t), population_i(t)]$.
$infected_i(t)$ is the number of newly reported infection cases, and $population_i(t)$ denotes the bird abundance in location $i$.
Consistent with empirical epidemiological findings emphasizing localized avian influenza transmission \citep{bonney2018spatial}, we then define spatial edges between locations using a kernel-based weighting scheme rather than a rigid distance cutoff. 
For each undirected spatial edge $e^{(sp)}_{<i,j>}$ connecting locations $i$ and $j$, the weight $\omega_{ij}^{sp}$ is assigned using a Gaussian kernel $\omega_{ij}^{sp} = K(D_{ij}) = \exp\left(-\frac{D_{ij}^2}{2\sigma^2}\right)$.
Here, $D_{ij}$ is the geographic distance. $\sigma = \tau_d / 3$ is a parameter defining the connection range, ensuring that closer nodes exhibit stronger connections in the spatial layer.

\textbf{Location Layer. }
The case nodes $\mathcal{V}^{(p)}_t=\{v^{(p)}_1, \cdots, v^{(p)}_M\}$ form a time-varying set representing $M$ reported infected samples at time $t$.
The feature vector $\mathbf{x}_m^{(p)}(t)$ represents the genetic profile of case $m$. It is computed as the average pairwise genetic distance to other cases, obtained via the Kimura 2-parameter (K80) model \citep{kimura1980} on aligned hemagglutinin (HA) sequences.
Genetic edges $\{e^{(ge)}_{<m,n>}\}$ connect case $n$ and case $m$ based on their genetic similarity, defining the case layer.
The bi-layer heterogeneous graph is completed by connecting each case node $v^{(p)}_m$ to its specific reported location node $v^{(c)}_i$ via an assignment edge $e^{(as)}_{<i,m>}$. This inter-layer connection effectively integrates the dynamic case layer and the static location layer, forming a unified structure.

\subsection{Cross-layer Smoothing Block}
Modeling AIV transmission requires capturing complex interactions between the case layer and the location layer. 
To handle the discrepancy among heterogeneous nodes and edges, we propose the cross-layer smoothing block to leverage local dependencies in the heterogeneous graph neighborhoods. 
It encourages coherent node representations among epidemiologically linked groups while preserving type-specific semantics of case and location nodes.

The smoothing module utilizes a mean-field approach that iteratively refines the node embeddings on the heterogeneous graph. Given an initial node embedding $\mathbf{x}_{v}^{(0)}$ for node $v$ and the three distinct edge types ($\mathcal{E}^{(sp)}, \mathcal{E}^{(ge)}, \mathcal{E}^{(as)}$), we perform $K$ rounds of relation-specific message passing.
The relation-specific message $m_{r}^{(k)}(v)$ from neighbors connected by relation $r$ is calculated as the normalized average of transformed neighbor embeddings
\begin{equation}
    \begin{aligned}
        m_{r}^{(k)}(v) = \frac{1}{|\mathcal{N}_r(v)|}\sum_{u \in \mathcal{N}_r(v)}\mathbf{W}_{r} x_{u}^{(k-1)},
    \end{aligned}
\vspace{-2mm}
\end{equation}
where $\mathcal{N}_r(v)$ denotes the immediate neighbors of node $v$ under relation $r$. The matrix $\mathbf{W}_{r}$ is a trainable parameter matrix that represents the strength of interactions between connected nodes under relation $r$, thereby adhering to the local Markov property.
These relation-specific messages, which reflect the smoothed neighbor information, are then aggregated across all relations $r$ and combined with a node type-specific bias $\mathbf{b}_{\tau{(v)}}$ (where $\tau(v) \in \{c, p\}$ denotes the node type):
\begin{equation}
    \begin{aligned}
        \mathbf{x}_{v}^{(k)} = \text{ReLU}\left(\sum_r m_{r}^{(k)}(v) + \mathbf{b}_{\tau{(v)}}\right).
    \end{aligned}
\vspace{-2mm}
\end{equation}
Here, we employ the ReLU activation function, ensuring that node embedding $\mathbf{x}_{v}^{(k)}$ is non-linearly influenced by the semantics of its neighbors through the respective relations.
Iteratively applying this update rule $K$ times mimics belief propagation, systematically spreading information across the graph structure while explicitly considering the distinct relational contexts. Therefore, by restricting propagation to immediate neighbors, applying learnable, relation-specific transformations ($\mathbf{W}_r$), and iteratively refining node representations, this approach effectively integrates key aspects of MRF inference into a differentiable graph-based framework.

\subsection{Information-preserving Fusion Graphs}
Heterogeneous graphs, characterized by diverse types of nodes and relationships, exhibit an inherently complex structure stemming from the simultaneous presence of distinct local and global relational dependencies. 
Converting a heterogeneous graph into a homogeneous form can significantly simplify the representation, however, it risks losing valuable information embedded in the multi-source relations.
To overcome this limitation, we propose the Fusion Graph method. This method transforms the original heterogeneous structure (comprising location and case nodes connected by multiple relational types) into a unified representation under a principle of spectral alignment (details provided in Section \ref{sec:spec}). 
Consequently, The resulting Fusion Graph maintains the original heterogeneous complexity while offering a simpler, more interpretable structure (The timestep $t$ is omitted for notational clarity).
\paragraph{Fusion Nodes.}
Fusion Nodes are aggregated location representations, systematically integrating neighbor information from the heterogeneous graph $\mathcal{G}_t$.The Fusion Node embedding $\mathbf{x}_{i}$ for location $v_i^{(c)}$ is generated via a three-step fusion:

\begin{equation}
    \begin{aligned}
    \mathbf{z}^{(c)}_{i}
    &= f_{1}\!\bigl([\mathbf{x}_{i}^{(c)} \,\|\, \mathbf{x}_{i}^{(c\_\mathrm{spatial})}]\bigr) && \text{(Location Features + Spatial Context)} \\
    \mathbf{z}^{(p)}_{i}
    &= f_{2}\!\bigl([\mathbf{x}_{i}^{(p)} \,\|\, \mathbf{x}_{i}^{(p\_\mathrm{genetic})}]\bigr) && \text{(Genetic Context)} \\
    \mathbf{x}_{i}
    &= f_{(m)}\!\bigl([\mathbf{z}^{(c)}_{i} \,\|\, \mathbf{z}^{(p)}_{i}]\bigr) && \text{(Final Fusion)}
    \end{aligned}
\end{equation}
Here, $f_{1}$, $f_{2}$, and $f_{(m)}$ are Multi-Layer Perceptrons (MLPs) with non-linear activation functions (shared across time steps), and $[\cdot \,\|\, \cdot]$ denotes vector concatenation.
The contextual features $\mathbf{x}_{i}^{(c\_\mathrm{spatial})}$ and $\mathbf{x}_{i}^{(p\_\mathrm{genetic})}$ are calculated by averaging neighbor embeddings:
$\mathbf{x}_{i}^{(c\_\mathrm{spatial})}$ aggregates features from neighboring locations $\mathcal{N}_{i}^{(\mathrm{sp})}$: $\mathbf{x}_{i}^{(c\_\mathrm{spatial})} = \frac{1}{\lvert \mathcal{N}_{i}^{(\mathrm{sp})} \rvert}\, \sum_{j \in \mathcal{N}_{i}^{(\mathrm{sp})}} \mathbf{x}_{j}^{(c)}$.
Genetic Context $\mathbf{x}_{i}^{(p\_\mathrm{genetic})}$ aggregates genetic features from all cases $\mathcal{C}_{i}$ within location $i$:$\mathbf{x}_{i}^{(p\_\mathrm{genetic})} = \frac{1}{\lvert \mathcal{C}_{i} \rvert}\, \sum_{k \in \mathcal{C}_{i}} \mathbf{x}_{k}^{(p)}$.
The MLPs $f_{1}, f_{2}, f_{(m)}$ integrate the location's base features, spatial context, aggregated case features, and genetic context into a coherent, final Fusion Node embedding $\mathbf{x}_{i}$.

\vspace{-2mm}
\paragraph{Fusion Edges.}
Once the Fusion Node embeddings ($\mathbf{x}_i$) are obtained, we proceed to construct edges to induce a coherent relational topology for the homogeneous Fusion Graph. 
Rather than resorting to an exhaustive pairwise check or a simple hard distance cutoff, we employ a learnable link prediction network augmented with Locality-Sensitive Hashing (LSH) to efficiently select edges and ensure computational tractability.
The link probability between any pair of fusion nodes $v_i$ and $v_j$ is defined by $p_{ij} = \sigma(W_l[\mathbf{x}_{i} \| \mathbf{x}_{j}]+b_l)$,
where $\sigma(\cdot)$ is the Sigmoid function, $[\cdot \| \cdot]$ is concatenation, and $\mathbf{W}_l, \mathbf{b}_l$ are the learnable network parameters.
To avoid the computationally prohibitive $O(N^2)$ enumeration of all pairs, we use LSH to generate a reduced set of candidate edges. LSH projects the embeddings onto $B$ random hyperplanes to generate binary codes, approximating cosine similarity in the embedding space\citep{charikar2002similarity}.
Each fusion node embedding $\mathbf{x}_{i}$ is converted into a $B$-bit binary code $\mathbf{h}_i$
\begin{equation}
    \begin{aligned}
        \mathbf{h}_i = [\text{sign}(\mathbf{r}_1^{\top} \mathbf{x}_{i}), \cdots, \text{sign}(\mathbf{r}_B^{\top} \mathbf{x}_{i})] \in \{0,1\}^B
    \end{aligned}
\end{equation}
where $\mathbf{r}_h$ are independent random projection vectors.
Nodes that share identical hash codes are grouped, and these within-bucket pairs $(i, j)$ form the initial candidate edge set. 
This method relies on the principle that similar vectors have a high probability of LSH collision.
If the number of exact-match candidates is below a predefined maximum $M_{\text{max}}$, the candidate set is augmented by selecting pairs whose Hamming distance between their codes does not exceed a threshold $\tau_h$. By grouping nodes via code matches, we significantly reduce the cost of exhaustive pairwise comparison to approximately $O(N+M_{\text{max}})$ operations, where $N=|\mathcal{V}^{(c)}|$is the size of location nodes and the case node set is integrated into the fusion node embeddings in the Fusion Graph.

Following, to seamlessly integrate heterogeneous relations into the Fusion Graph, \ourmethod~employs a gating network that dynamically weights the influence of spatial and genetic edges based on the specific node-pair interactions.
For fusion nodes $v_i$ and $v_j$, we first compute the relation-specific embedding $\mathbf{e}_{ij}^{(r)} = \mathbf{e}_{r} + \mathbf{W}_{\text{edge}}\mathbf{x}_{ij}^{(r)} + \mathbf{b}_{\text{edge}}$ ($r \in \{\text{spatial}, \text{genetic}\}$).
These relation-specific embeddings are then fed into a multi-head self-attention network \citep{vaswani2017attention} to produce normalized scores $\alpha_{ij}^{(r)}$ across the relations, and the final fusion-edge embedding $\mathbf{e}_{ij}$ is then obtained as a weighted sum of the relation-specific embeddings
\vspace{-2mm}
\begin{equation}
    \mathbf{\alpha}_{ij}^{(r)} = \text{Attn}(\mathbf{x}_{i}, \mathbf{x}_{j}, \mathbf{e}_{ij}^{(r)}), \quad
    \mathbf{e}_{ij} = \sum_{r} \alpha_{ij}^{(r)}\mathbf{e}_{ij}^{(r)}
\vspace{-1mm}
\end{equation}
This process yields the unified Fusion Graph characterized by the updated fusion node embeddings $\mathbf{X}_t$ and the aggregated edge embeddings $\mathbf{E}^{(f)}_t$.
To ensure maximum information preservation during the fusion process, we employ a spectral regularizer defined by the Frobenius norm of the difference between the Laplacian matrices of the two graph structures: $\mathcal{L}_{\text{spectral}} = \bigl\|\,\mathbf{L}_{\text{hetero}} - \mathbf{L}_f \bigr\|_{F}^{2}$.
It encourages consistency in the diffusion modes (eigenvectors and eigenvalues) between the original heterogeneous graph's Laplacian ($\mathbf{L}_{\text{hetero}}$) and the Fusion Graph's Laplacian ($\mathbf{L}_f$), thereby preserving the graph's global structural geometry (detailed in Section \ref{sec:spec}).

\vspace{-2mm}
\subsection{Autoregressive Forecasting}
To effectively model temporal dependencies, we adopt a sequence-to-sequence architecture operating over the compressed Fusion Graphs, using an encoder for feature extraction and an autoregressive decoder for forecasting.
The encoder processes the Fusion Graph sequence using $L$ layers of GNN ($\text{G}_l$) at each time step $t$
\begin{equation}
    \begin{aligned}
           \mathbf{H}_t^{(l+1)} = \sigma \left( \text{G}_{l}(\mathbf{H}_t^{(l)}, \mathbf{E}^{(f)}_t) \right) \quad \text{for } l=0, 1, \cdots, L-1
    \end{aligned}
\end{equation}
The initial layer input is set as $\mathbf{H}_t^{(0)} = [\mathbf{X}_t + \mathbf{p}_t]$, where $\mathbf{X}_t$ contains the fusion node embeddings, and $\mathbf{p}_t$ represents a learnable positional encoding to capture the temporal order. 
Over the $T$ observed timesteps, the final layer outputs are collected as $\mathcal{H} \in \mathbb{R}^{w \times N \times d} = \{\mathbf{H}_T^{(L)}, \cdots, \mathbf{H}_{T+w-1}^{(L)}\}$. 
This tensor is then input into a multi-head attention network \citep{vaswani2017attention}, enforcing mutual interaction across features from different timesteps and yielding a global context vector $\mathbf{H}^{(c)}$.

The decoder operates in an autoregressive manner over a forecasting horizon $H$. 
At each decoding step $h \in \{1, \cdots, H\}$, it employs a GraphSAGE-based architecture (comprising $L$ layers mirroring the encoder's structure) to produce the decoded features $\mathbf{d}_h$
\begin{equation}
    \begin{aligned}
    \mathbf{d}_h = \text{Decoder}(\mathbf{Z}_{h-1}^{(L)}, \mathbf{E}_{T}^{(f)}, 
    \end{aligned}
\end{equation}
where $\mathbf{Z}_{h-1}^{(L)}$ is the hidden state (initialized with $\mathbf{Z}_{0}^{(L)}=\mathbf{H}^{(c)}$). 
$\mathbf{E}^{(f)}_T$ uses the Fusion Edges from the last observation $T$, and $\mathbf{t}_h$ is a temporal encoding. 
The decoded output from the previous step is incorporated via a linearly projected feature vector $\mathbf{f}_{h-1} = \mathbf{W}_{\text{feat}} \cdot \hat{\mathbf{y}}_{h-1} + \mathbf{b}_{\text{feat}}$.
Here, $\hat{\mathbf{y}}_{h-1}$ is the predicted outcome from the prior step, used as input for the current step.
The decoded features $\mathbf{d}_h$ are then integrated with the prior prediction $\mathbf{f}_{h-1}$ and the global context $\mathbf{H}^{(c)}$ through a linear combination
\begin{equation}
    \tilde{\mathbf{d}}_h = (1- \lambda_o -\lambda_p) \mathbf{d}_h + \lambda_o \mathbf{f}_{h-1} + \lambda_p \mathbf{H}^{(c)}
\end{equation}
The parameters $\lambda_o, \lambda_p$ control the reliance on the current decoding, prior prediction, and global context, respectively, enhancing forecast stability.
Finally, the prediction for step $h$, $\hat{\mathbf{y}}_h$, is generated via a non-linear projection $\hat{\mathbf{y}}_h = \mathbf{W}_{\text{out}}\tilde{\mathbf{d}}_h$.

\paragraph{Spectral Alignment}
\label{sec:spec}
Conceptually, compressing the bi-layer graph into the Fusion Graph acts as a spectral low-pass filter.
It discards the high-frequency components associated with the fine-grained, case-level sub-graph (e.g., specific individual genetic links).
Consequently, operating solely on the fused graph risks subtle transmission channels that are often driven by a few genetically distinctive samples.
To mitigate this information loss and optimally preserve the epidemiologically useful structural information, we employ a spectral regularizer $\| \mathbf{L}_{\text{hetero}} - \tilde{\mathbf{L}}_f \|$. 
It enforces \textbf{spectral alignment} between between the Laplacian matrices of the original bi-layer heterogeneous graph and the resulting homogeneous Fusion Graph, ensuring consistency in their global diffusion properties.

Let $P \in \{0,1\}^{(N+M_{\text{max}})\times N}$ be the projection matrix mapping case nodes to their home locations.
$\mathbf{L}_{\text{hetero}}$ and $\mathbf{L}_f$ be the Laplacians of the heterogeneous and fusion graphs, respectively. Define the projected fusion Laplacian $\tilde{\mathbf{L}}_f:=P\mathbf{L}_fP^\top$, and the diffusion operators $\mathbf{M}_{\text{hetero}}=\mathbf{I} - \mathbf{L}_{\text{hetero}}$ and $\tilde{\mathbf{M}}_f = \mathbf{I}-\tilde{\mathbf{L}}_f$.
\begin{theorem}
\label{tm}
    Assuming the spectral approximation error is bounded by $\| \mathbf{L}_{\text{hetero}} - \tilde{\mathbf{L}}_f \| \leq \varepsilon$, for any polynomial filter $p(\cdot)$ and feature matrix $\mathbf{H}$, the difference between applying the heterogeneous and projected fusion operators is bounded:
    \begin{equation}
        \| p(\mathbf{M}_{\text{hetero}})\mathbf{H} - p(\tilde{\mathbf{M}}_{f})\mathbf{H} \|_{F} \leq \mathcal{O}(\varepsilon)\|\mathbf{H}\|_F
    \end{equation}
\end{theorem}
The theorem demonstrates that if the Laplacians are spectrally close (difference bounded by $\varepsilon$), the application of standard polynomial GNN filters to the node features $\mathbf{H}$ results in a bounded error $\mathcal{O}(\varepsilon)$ between the heterogeneous and fused representations. This is a foundational result ensuring that the Fusion Graph can approximate the original graph's spectral operations.

Applying Theorem \ref{tm}, we analyze the spectral consistency of a single GraphSAGE layer, $\Phi(\mathbf{H})=\sigma(\mathbf{W}_s\mathbf{H} + \mathbf{W}_n p(\mathbf{M}) \mathbf{H}),$ where $\mathbf{W}_s$ and $\mathbf{W}_n$ are weights with norms $\|\mathbf{W}_s\|\leq \beta_s$ and $\|\mathbf{W}_n\|\leq \beta_n$.
Defining the Lipschitz constant $Z:=\beta_s + \beta_n \| p(\mathbf{M}_{\text{hetero}}) \|_2$, the single-layer error bound is
\begin{equation}
    \|\Phi_{\text{hetero}}(\mathbf{H}) - \Phi_f(P^\top \mathbf{H})\|_F \leq \mathcal{O}(\beta_n \varepsilon) \|\mathbf{H}\|_F + \underbrace{\beta_s\| \mathbf{H} - P P^\top \mathbf{H} \|_F}_{\text{fusion mismatch term}}.
\end{equation}
Since the heterogeneous node set $\mathbf{H}$ is fused at the location level ($\mathbf{H} = P P^\top \mathbf{H}$), the fusion mismatch term simplifies to zero, yielding
\begin{equation}
    \|\Phi_{\text{hetero}}(\mathbf{H}) - \Phi_f(P^\top \mathbf{H})\|_F \leq \mathcal{O}(\beta_n \varepsilon) \|\mathbf{H}\|_F.
\end{equation}
For \ourmethod~implemented with $L$ stacked GraphSAGEs (denoted as $\mathcal{F}$), the accumulated error is bounded by a geometric series
\begin{equation}
    \| \mathcal{F}_{\text{hetero}}(\mathbf{H}) - \mathcal{F}_f(P^{\top}\mathbf{H}) \|_F \leq \mathcal{O}\left(\frac{Z^{L}-1}{Z-1} \varepsilon\right) \|\mathbf{H}\|_F.
\end{equation}
The theoretical bound demonstrates that the accumulated error grows geometrically with the network depth $L$, governed by the $Z$.
To achieve a tighter and well-behaved upper bound, we apply weight normalization to the GraphSAGE layers $\Phi(\mathbf{H})$, ensuring that $Z < 1$.
Consequently, the GraphSAGE upper error bound, which converges to a constant value $\mathcal{O}\left(\frac{1}{1-Z} \varepsilon\right)$, is theoretically lower and more stable than the $\mathcal{O}(L\varepsilon)$ linear bound for any depth $L \ge 2$ (only performs linear bound when $L = 1$).

\vspace{-2mm}
\paragraph{Theoretical justification.}  
We selected GraphSAGE as the backbone as it ensures the theoretical tighter upper bound of $\mathcal{L}_{\text{spectral}}$.
As mentioned, \ourmethod~relies on the spectral regularizer $\mathcal{L}_{\text{spectral}}$ to ensure the Fusion Graph preserves the original heterogeneous graph's diffusion geometry.
The suitability of candidate GNN architectures, GraphSAGE, GCN, and GAT, is as follows:
\begin{itemize}
\item \textbf{GCN:} GCN uses repeated application of the symmetric normalized Laplacian matrix $\mathbf{D}^{-1/2} \mathbf{A} \mathbf{D}^{-1/2}$.When this operator deviates from the target Laplacian by an error of $\varepsilon$, the cumulative spectral deviation over $L$ layers accumulates linearly, resulting in an error up to $\mathcal{O}(L\varepsilon)$. This linear error accumulation can rapidly degrade the spectral alignment. 

\item \textbf{GraphSAGE:} Its inductive neighborhood aggregation provides a tighter error bound. By explicitly regularizing the Lipschitz constant $Z < 1$, the cumulative error is bounded by $\mathcal{O}\left(\frac{Z^{L}-1}{Z-1} \right) \varepsilon$. This bound is constantly lower than the GCN bound $\mathcal{O}(L\varepsilon)$ and ensures the cumulative error remains controlled and stable, making GraphSAGE theoretically well-aligned with the spectral regularization objective.

\item \textbf{GAT:} GAT dynamically adjusts the adjacency via feature-dependent attention. The resulting propagation operator is not fixed or governed by the regularized Laplacian $\mathbf{L}_f$. This evolving topology makes spectral consistency guarantees, which require a stable operator, inapplicable.
\end{itemize}

\vspace{-3mm}
\subsection{Optimization}
\label{sec:opt}
\ourmethod~is trained by minimizing a total objective function composed of three components:
(i) the multi-step forecasting term ($\mathcal{L}_{\text{pred}}$),
(ii) the spectral alignment regularizer ($\mathcal{L}_{\text{spectral}}$), and
(iii) parameter regularization ($\mathcal{L}_{\text{reg}}$).
The forecasting term measures the error between the forecast $\hat{y}_{i,h}$ and the ground-truth count $y_{i,h}$. Because epidemiological requirements prioritize accurate prediction of high infection counts, we employ a hierarchical weighting scheme based on the magnitude of actual infections
\vspace{-1mm}
\begin{equation}
    \begin{aligned}
        \mathcal{L}_{pred} = \frac{1}{N} \sum_{h=T+1}^{T+H}\sum_{i=1}^{N} w_i \cdot (\hat{y}_{i,h} - y_{i,h})^2,
    \end{aligned}
\vspace{-1mm}
\end{equation}
Here, $N$ is the number of locations, $H$ is the forecast horizon length, and $w_i$ is a weight coefficient based on the infection severity $\tau_i \in \{\text{low}, \text{medium}, \text{high}\}$ of location $i$. By assigning higher weights to regions with higher infection counts, \ourmethod~concentrates its learning efforts on non-zero infections through this progressive weighting, enhancing its predictive accuracy on critical outbreak scenarios.

With $\lambda_1, \lambda_2$ controlling the weights of spectral alignment and regularization, the final objective is
\vspace{-2mm}
\begin{equation}
    \mathcal{L}_{\mathrm{tot}}
     \;=\;
     \mathcal{L}_{\mathrm{pred}}
     \;+\;
     \lambda_{\mathrm{1}}\,\cdot
     \bigl\|\,\mathbf{L}_{hetero} - \tilde{\mathbf{L}}_f \bigr\|_{F}^{2}
     \;+\;
     \lambda_{2}\,
     \sum_{w\in\Theta}\|w\|_{2}^{2}.
\vspace{-2mm}
\end{equation}

\vspace{-3mm}
\section{Experiments}
\label{sec:exp}
We evaluate \ourmethod~ using our newly released \textbf{Avian-US} dataset\footnote{\url{https://huggingface.co/datasets/CRUISEResearchGroup/Avian-US_dataset}} (detailed in Appendix \ref{apd:dataset}), which incorporates genetic, spatial, and ecological data across 3,227 counties in the United States. 
The goal is to empirically demonstrate \ourmethod's ability to effectively leverage geographical and genomic information to model multi-scale transmission dynamics.

\vspace{-2mm}
\paragraph{Baselines.}
\label{sec:baselines}
We compare \ourmethod~against:
1) homogeneous spatiotemporal models (\textbf{STGCN}~\citep{yu2018spatio}, \textbf{SelfAttnRNN}~\citep{cheng2016long}, \textbf{DCRNN}~\citep{li2018diffusion}, \textbf{EAST-Net}~\citep{wang2022event} with a simplify version \textbf{ST-Net}), 2) epidemic prediction models (\textbf{Cola-GNN}~\citep{deng2020cola}, \textbf{EpiGNN}~\citep{xie2022epignn}, \textbf{Epi-Cola-GNN}~\citep{liu2023epidemiology}, \textbf{STSGT}~\citep{banerjee2022spatial}), and 3) heterogeneous GNN model (\textbf{HGT}~\citep{hu2020heterogeneous}).

\vspace{-2mm}
\paragraph{Experimental settings.}
\label{sec:setting}
We adopt 5 complementary metrics: Root Mean Squared Error (\textbf{RMSE}), Mean Absolute Error (\textbf{MAE}), Pearson Correlation Coefficient (\textbf{PCC}), Spearman Correlation Coefficient (\textbf{SCC}), and Threshold F1 Score (\textbf{F1}, only detected when prediction exceeds threshold $t=0.3$).
Please refer to Appendix \ref{apd:setting} for implementation details and experimental settings.
We set a short observation window of $T$=4 steps and forecast the next $H$=4 steps (all experiments are conducted under this setting unless specified), and report the averaged evaluation metrics of $H$ steps. 
Experiments of all baselines and BLUE are conducted under a 5-fold cross-validation with the same random seed to ensure consistency. 
For detailed hyperparameter setting, please refer to Appendix \ref{apd:setting}.

\begin{table*}[]
\caption{Overall Performance. Experiments are run under $T$=4 and $H$=4. * denote statistically significant improvements, validated by a paired t-test at a significance level of $p < 0.05$ against the \underline{runner-up model}. Best results are \textbf{bolded}.}
\vspace{-3mm}
\label{tab:overall}
\resizebox{0.9\linewidth}{!}{
    \begin{tabular}{l|ccccc}
    \toprule
    Model         & RMSE($\downarrow$)       & MAE($\downarrow$)        & PCC($\uparrow$)          & SCC($\uparrow$)          & F1($\uparrow$)           \\ \hline
    STGCN         & 0.8741±0.1428            & 0.4198±0.0911            & 0.0481±0.0024            & 0.0773±0.0087            & 0.0637±0.0043            \\
    SelfAttnRNN   & 0.8962±0.1752            & 0.3722±0.0566            & 0.0523±0.0052            & 0.0800±0.0111            & 0.0698±0.0038            \\
    ST-Net        & 0.9020±0.1603            & 0.4572±0.0840            & 0.0584±0.0034            & 0.0865±0.0128            & 0.0751±0.0167            \\
    EAST-Net      & 0.8973±0.1772            & 0.5556±0.0626            & 0.0647±0.0037            & 0.0839±0.0134            & 0.0779±0.0083            \\
    DRCNN         & 0.7965±0.1314            & 0.6123±0.0395            & 0.0588±0.0053            & 0.0871±0.0117            & 0.0665±0.0078            \\ \hline
    EpiGNN        & 0.7834±0.1200            & 0.2627±0.0440            & 0.0719±0.0049            & 0.0816±0.0112            & 0.0672±0.0053            \\
    Cola-GNN      & 0.7118±0.1001            & 0.1777±0.0223            & 0.0770±0.0042            & 0.0846±0.0102            & 0.0653±0.0096            \\
    Epi-Cola-GNN  & 0.8264±0.1246            & 0.0967±0.0172            & {\ul 0.0780±0.0042}      & 0.0825±0.0112            & {\ul 0.0816±0.0197}      \\
    STSGT         & 0.6907±0.1143            & 0.1690±0.0312            & 0.0403±0.0080            & {\ul 0.1082±0.0201}      & 0.0725±0.0183            \\ \hline
    HGT           & {\ul 0.6523±0.0828}      & {\ul 0.0902±0.0163}      & 0.0772±0.0054            & 0.0966±0.0165            & 0.0801±0.0077            \\ \hline
    \textbf{BLUE} & \textbf{0.6106 ± 0.0719} & \textbf{0.0848 ± 0.0182} & \textbf{0.0871 ± 0.0132} & \textbf{0.1217 ± 0.0140} & \textbf{0.1001 ± 0.0193} \\ \bottomrule
    \end{tabular}
    }
    \vspace{-2mm}
\end{table*}

\vspace{-2mm}
\subsection{Overall Performance}
\ourmethod~achieves the highest PCC. 
However, since PCC is sensitive to small fluctuations and can be skewed in sparse datasets, we also incorporate the SCC, a rank-based metric suitable for capturing sparse, non-linear epidemic signals.
As shown in Table.\ref{tab:overall}, HGT is the runner-up in regression-based metrics (RMSE/MAE) due to its capacity to model heterogeneous nodes.
Epi-Cola-GNN excels at capturing linear correlations, while STSGT demonstrates superior performance in outbreak detection, achieving the runner-up F1 score.
\ourmethod~achieves the highest PCC and SCC, confirming its strong capacity to capture both linear and non-linear correlations inherent in the AIV transmission data.
Across all metrics, \ourmethod~exhibits a clear advantage: it reduces the RMSE and MSE by 0.0054 compared to the runner-up (HGT) and provides reasonable improvements in PCC and SCC against all baselines. 
Furthermore, \ourmethod~achieves the highest F1 score, which highlights its superior ability to detect outbreak occurrences beyond merely predicting case counts.

\begin{figure}
    \centering
    
    \begin{subfigure}[b]{.95\linewidth}
        \centering
        \includegraphics[width=\textwidth]{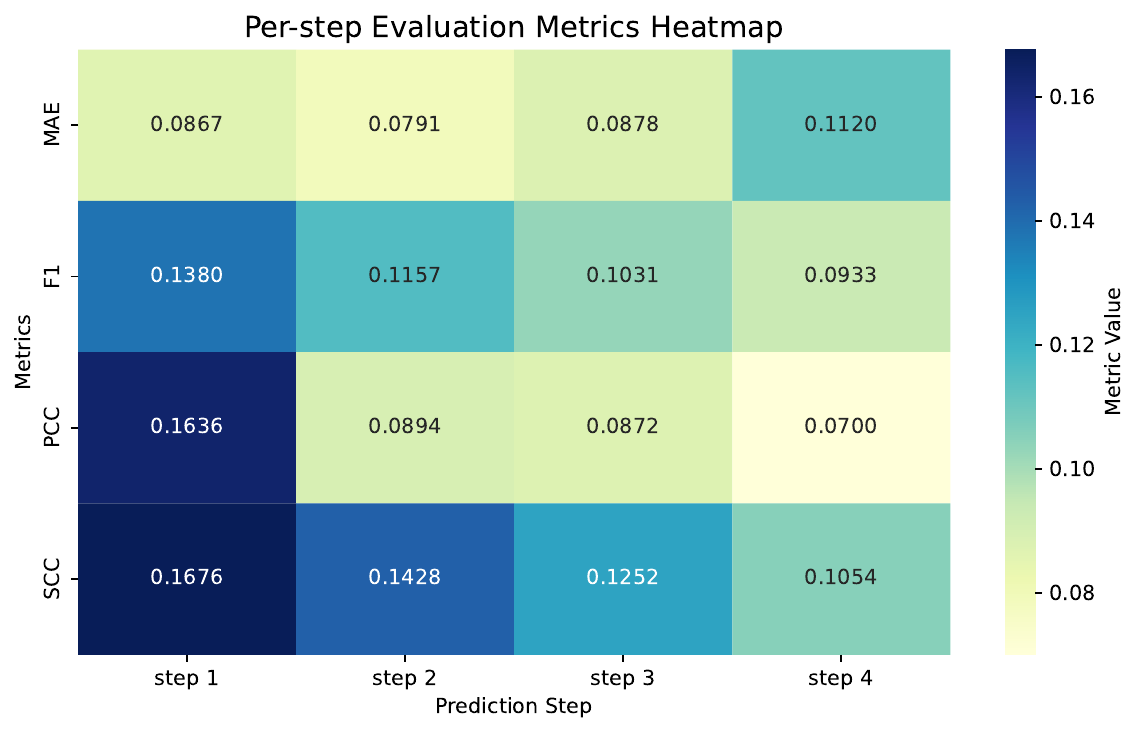}
    \end{subfigure}
    \begin{subfigure}[b]{.95\linewidth}
        \centering
        \includegraphics[width=\textwidth]{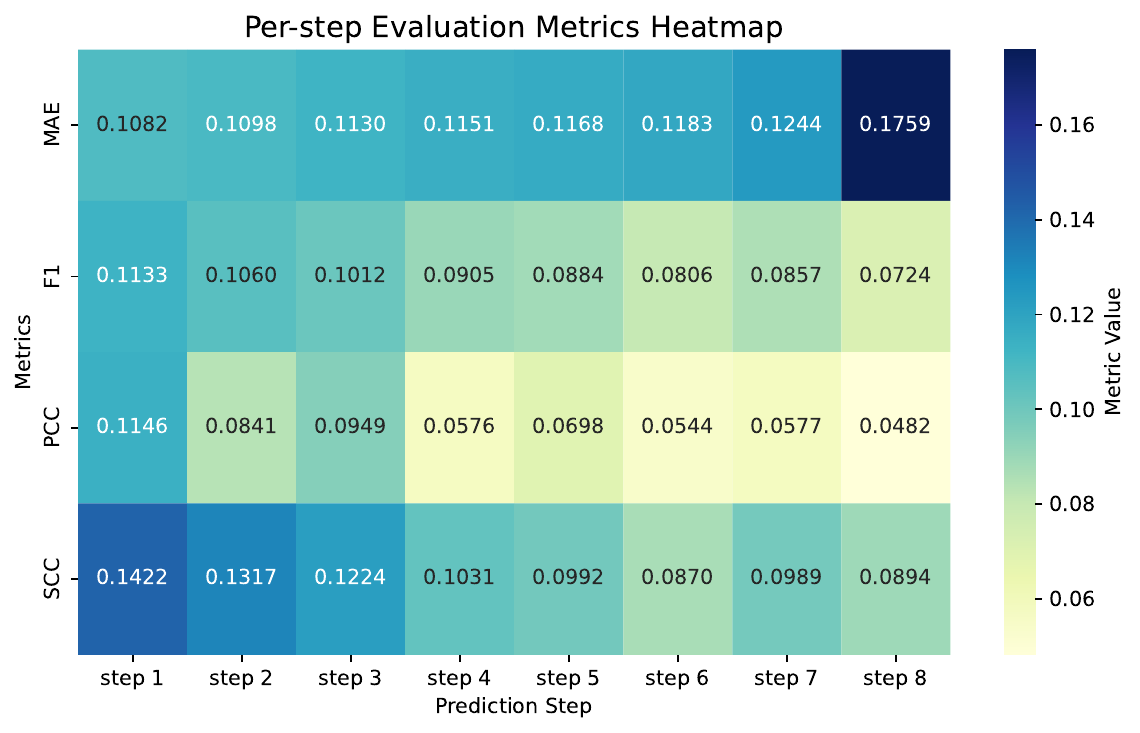}
    \end{subfigure}
    \vspace{-4mm}
    \caption{Per-step performance on Avian-US with $H=4$ (up) and $H=8$ (down).}
    \label{fig:perstep}
\end{figure}

To provide finer-grained insights into forecasting performance, we conducted a temporal analysis of per-step results under a look-back window $T=4$ and forecasting horizons $H \in \{4, 8\}$.
Consistent with the challenge of long-range forecasting, nearly all evaluation metrics degrade as the prediction horizon $H$ increases, shown in Fig. \ref{fig:perstep}.
\ourmethod~performs best at Step 1 for both $H=4$ and $H=8$, achieving its peak F1 score, PCC, and SCC.
Performance then consistently declines as the predicted step increases, resulting in the lowest outbreak detection ability and largest prediction errors at the final step. 
Specifically, \ourmethod's capacity to detect outbreaks decreases, with the F1 score falling from 0.1380 to 0.0933 for $H=4$ and from 0.1133 to 0.0724 for $H=8$.
SCC also shows a consistent decrease.
Despite slight fluctuations, MAE of predicted infection counts increases with each successive step, rising from 0.0867 to 0.1120 for $H=4$ and from 0.1082 to 0.1759 for $H=8$.

\vspace{-2mm}
\subsection{Ablation Study}

\begin{table}
  \centering
    \centering
    \vspace{-2mm}
    \caption{Ablation study on Avian-US.}
    \vspace{-3mm}
    \label{tab:ablation}
    \resizebox{.95\linewidth}{!}{
    \begin{tabular}{l|ccccc}
    \toprule
    Variants    & RMSE            & MAE             & F1              & PCC             & SCC              \\ \hline
    w/o gen     & 0.8093          & 0.1667          & 0.0721          & 0.0686          & 0.0946          \\
    w/o CS      & 0.7824          & 0.1828          & 0.0692          & 0.0677          & 0.0911          \\
    w/o Spec    & 0.8504          & 0.2310          & 0.0839          & 0.0729          & 0.1035          \\
    w/o CS+Spec & 0.9020          & 0.2002          & 0.0605          & 0.0543          & 0.0869          \\
    w/ LSH      & 0.6141          & 0.1007          & 0.0994          & 0.0855          & 0.1105            \\
    w/ attn     & 0.6157          & 0.0945          & 0.0901          & 0.0755          & 0.1114            \\
    w/o eco     & 0.6772          & 0.1373          & 0.0978          & 0.0771          & 0.1093          \\
    w/ drop     & 0.6713          & 0.1254          & 0.0944          & 0.0801          & 0.1153          \\ \hline
    BLUE        & \textbf{0.6106} & \textbf{0.0848} & \textbf{0.1020} & \textbf{0.0871} & \textbf{0.1218}          \\ \bottomrule
    \end{tabular}
    }
    \vspace{-3mm}
\end{table}

We validate the design choices in \ourmethod~by comparing \ourmethod against eight variants:
1) \textit{w/o gen} only utilizes the spatial distances and assignment relationships;
2) \textit{w/o CS+Spec} only implement the auto-regressive encoder-decoder framework;
3) \textit{w/o CS} excludes the cross-layer smoothing block;
4) \textit{w/o Spec} excludes the spectral regularizer from the objective function;
5) \textit{w/ LSH} removes attention gate with simple averaging;
6) \textit{w/ attn} only use attention gate for fusion edge calculation;
7) \textit{w/o eco} removes location ecological features (bird abundance);
8) \textit{w/ drop} randomly drop 20\% of edges from bi-layer heterogeneous graphs.
The results are shown in Table. \ref{tab:ablation}.

The CS block and the Spectral Regularizer function work collaboratively. 
Disabling CS (\textit{w/o CS}) allows noise from the initial sparse graph construction to propagate, harming ranking metrics. 
Similarly, removing the spectral constraint (\textit{w/o Spec}) decouples the learned graph structure from the true epidemiological diffusion process, leading to overfitting.
The combined removal of these components (\textit{w/o CS+Spec}) yields the poorest performance, demonstrating that raw auto-regressive encoding is insufficient without structurally guided regularization.
Removing genetic edges (\textit{w/o gen}) causes a sharp increase in MAE, confirming that AIV outbreaks follow complex biological pathways that spatial proximity cannot fully explain.
This is further supported by the \textit{w/ drop} experiment, where randomly severing $20\%$ of edges degrades performance, showing that AIV spread forecasting relies on a reliable, complete connectivity structure for aggregation.
Moreover, two key observations arise from the attention experiments. 
First, replacing the dynamic gate with simple averaging (\textit{w/ LSH}) increases MAE, proving that the model must learn to weigh spatial versus genetic risks adaptively. 
Second, using full dense attention (\textit{w/ attn}) underperforms compared to the LSH-based approach (reducing F1 to $0.0901$). This suggests that LSH acts as a beneficial constraint—limiting aggregation to high-probability neighborhoods prevents the model from overfitting to noisy, long-range interactions common in sparse data.
Overall, \ourmethod~achieves the best performance across all metrics, confirming that each component contributes distinctly to capturing the spatiotemporal dynamics of AIV spread.

\vspace{-3mm}
\subsection{Spectral Alignment $\mathcal{L}_{spec}$}
\begin{figure}
\centering
    \begin{subfigure}[b]{.32\linewidth}
        \centering
        \includegraphics[width=\textwidth]{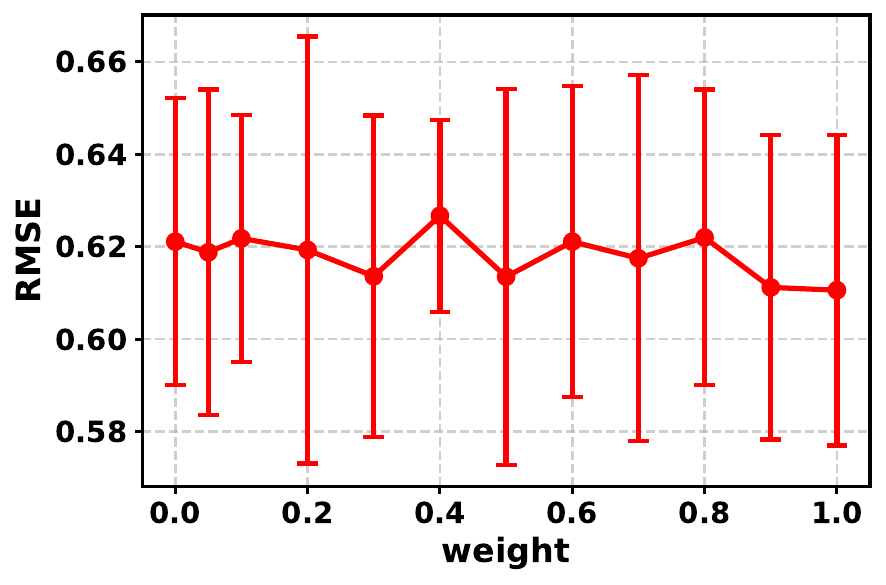}
        \label{fig：rmse}
    \end{subfigure}
    \begin{subfigure}[b]{.32\linewidth}
        \centering
        \includegraphics[width=\textwidth]{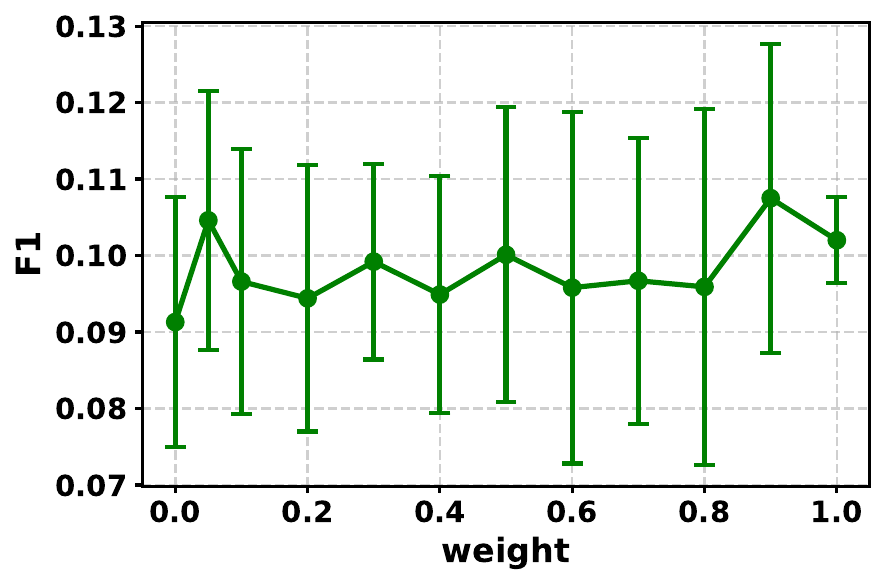}
        \label{fig:f1}
    \end{subfigure}
    \begin{subfigure}[b]{.32\linewidth}
        \centering
        \includegraphics[width=\textwidth]{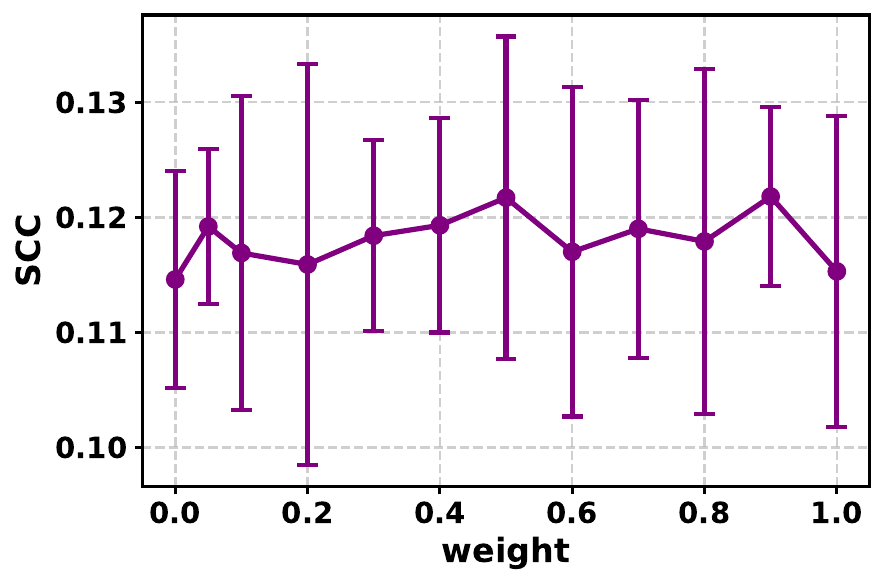}
        \label{fig:scc}
    \end{subfigure}
    \vspace{-5mm}
    \caption{Impact of spectral alignment weight $\lambda_1$.}
    \vspace{-5mm}
    \label{fig: lambda_1}
\end{figure}

We analyze the impact of the spectral regularizer weight, $\lambda_1$, which controls the trade-off between structural transmission alignment and forecasting accuracy. 
As shown in Figure \ref{fig: lambda_1}, incorporating $\mathcal{L}_{spec}$ significantly boosts performance across all metrics.
Specifically, as $\lambda_1$ is tuned from $0$ to $0.9$, we observe a consistent reduction in forecasting error, with RMSE dropping from $0.6211$ to $0.6112$. 
The model's ability to identify outbreak events improves drastically, with the F1 score peaking at $\lambda_1=0.9$, a $17.1\%$ increase compared to the baseline ($\lambda_1=0$). 
SCC follows an identical trend, confirming that spectral constraints assist the model in capturing correct correlation patterns.
It is worth noting that at $\lambda_1=1.0$, performance slightly declines, though the error bars narrow, indicating higher stability. 
This suggests that while spectral regularization is beneficial, an excessively large $\lambda_1$ forces the model to over-prioritize structural alignment at the expense of the main forecasting objective.

\vspace{-2mm}
\subsection{Backbone Choice}



We validate our architectural choice by replacing GraphSAGE with GCN and GAT while maintaining fixed experimental settings ($L=2$). 
Note that the spectral regularizer ($\lambda_1 = 0.2$) was excluded for GAT as its attention mechanism conflicts with the static spectral constraints.
Empirical results confirm that GraphSAGE consistently exhibits superior representational capability within our framework, outperforming alternatives in RMSE, MAE, and F1 scores.
Architecturally, GraphSAGE’s inductive neighbor sampling aligns more effectively with \ourmethod's information-preserving design than GCN or GAT. 
Consequently, GraphSAGE is retained as the optimal encoder for capturing the spatiotemporal dynamics of AIV spread.

\begin{table}[h]
\centering
\vspace{-2mm}
\caption{Performance comparison.}
\vspace{-2mm}
\begin{tabular}{l|ccc}
\toprule
Metrics & \textsc{BLUE} & w/GCN & w/GAT \\
\midrule
RMSE & \textbf{0.6106} & 0.6678 & 0.6791 \\
MAE  & \textbf{0.0848} & 0.0970 & 0.0942 \\
F1   & \textbf{0.1001} & 0.0967 & 0.0893 \\
PCC  & \textbf{0.0871} & 0.0664 & 0.0696 \\
SCC  & \textbf{0.1217} & 0.0931 & 0.1089 \\
\bottomrule
\end{tabular}
\vspace{-3mm}
\end{table}

\vspace{-2mm}
\section{Conclusion and Discussion}
\label{sec:discuss}
In this work, we proposed \ourmethod, a bi-layer heterogeneous graph fusion framework designed to advance epidemic forecasting by integrating genetic and spatial modalities. 
\ourmethod~uses cross-layer smoothing and information-preserving graph fusion to learn coherent representations of disease spread through an autoregressive encoder–decoder architecture.
Through a novel integration of cross-layer smoothing and spectral regularization, the model effectively learns coherent representations of disease transmission dynamics within an auto-regressive encoder-decoder architecture. 
Extensive empirical validation on the newly constructed Avian-US benchmark demonstrates that \ourmethod~consistently outperforms state-of-the-art spatiotemporal and epidemic forecasting baselines, establishing its robustness across diverse spatial and epidemiological settings.
Beyond immediate performance gains, \ourmethod~offers a generalizable theoretical contribution, providing a mathematically grounded method for maintaining structural alignment in complex, multi-view graph learning.
Although the current study relies on AIV, the architecture is inherently extensible.
Future iterations will exploit this extensibility to incorporate environmental covariates and broaden the biological scope to include cross-species transmission, extending \ourmethod~to comprehensive real-world AIV monitoring.

\vspace{-2mm}
\section{Acknowledgment}
This research is supported by the Australian Commonwealth Scientific and Industrial Research Organisation (CSIRO) and the United States National Science Foundation under Grant No. 2302968, titled "Understanding Bias in AI Models for the Prediction of Infectious Disease Spread". 
This research is conducted by the ARC Centre of Excellence for Automated Decision-Making and Society (No. CE200100005), funded by the Australian Government through the Australian Research Council.
We acknowledge the utilization of computational resources from the Katana High Performance Computing (HPC) cluster, which is supported by the Faculty of Engineering, UNSW Sydney. 
We also acknowledge the National Computational Infrastructure (NCI) for providing access to the Gadi supercomputer.

\clearpage
\bibliographystyle{ACM-Reference-Format}
\bibliography{arxiv}

@article{geng2021kernel,
  title={A kernel-modulated SIR model for Covid-19 contagious spread from county to continent},
  author={Geng, Xiaolong and Katul, Gabriel G and Gerges, Firas and Bou-Zeid, Elie and Nassif, Hani and Boufadel, Michel C},
  journal={Proceedings of the National Academy of Sciences},
  volume={118},
  number={21},
  pages={e2023321118},
  year={2021},
  publisher={National Academy of Sciences}
}

@article{della2023sir,
  title={An SIR model with viral load-dependent transmission},
  author={Della Marca, Rossella and Loy, Nadia and Tosin, Andrea},
  journal={Journal of Mathematical Biology},
  volume={86},
  number={4},
  pages={61},
  year={2023},
  publisher={Springer}
}

@inproceedings{liu2023human,
  title={Human mobility modeling during the COVID-19 pandemic via deep graph diffusion infomax},
  author={Liu, Yang and Rong, Yu and Guo, Zhuoning and Chen, Nuo and Xu, Tingyang and Tsung, Fugee and Li, Jia},
  booktitle={Proceedings of the AAAI Conference on Artificial Intelligence},
  volume={37},
  number={12},
  pages={14347--14355},
  year={2023}
}

@inproceedings{tang2023enhancing,
  title={Enhancing spatial spread prediction of infectious diseases through integrating multi-scale human mobility dynamics},
  author={Tang, Yinzhou and Wang, Huandong and Li, Yong},
  booktitle={Proceedings of the 31st ACM International Conference on Advances in Geographic Information Systems},
  pages={1--12},
  year={2023}
}

@inproceedings{liu2024review,
  title={A review of graph neural networks in epidemic modeling},
  author={Liu, Zewen and Wan, Guancheng and Prakash, B Aditya and Lau, Max SY and Jin, Wei},
  booktitle={Proceedings of the 30th ACM SIGKDD Conference on Knowledge Discovery and Data Mining},
  pages={6577--6587},
  year={2024}
}

@article{bonney2018spatial,
  title={Spatial transmission of H5N2 highly pathogenic avian influenza between Minnesota poultry premises during the 2015 outbreak},
  author={Bonney, Peter J and Malladi, Sasidhar and Boender, Gert Jan and Weaver, J Todd and Ssematimba, Amos and Halvorson, David A and Cardona, Carol J},
  journal={PloS one},
  volume={13},
  number={9},
  pages={e0204262},
  year={2018},
  publisher={Public Library of Science San Francisco, CA USA}
}

@inproceedings{charikar2002similarity,
  title={Similarity estimation techniques from rounding algorithms},
  author={Charikar, Moses S},
  booktitle={Proceedings of the thiry-fourth annual ACM symposium on Theory of computing},
  pages={380--388},
  year={2002}
}

@article{vaswani2017attention,
  title={Attention is all you need},
  author={Vaswani, Ashish and Shazeer, Noam and Parmar, Niki and Uszkoreit, Jakob and Jones, Llion and Gomez, Aidan N and Kaiser, {\L}ukasz and Polosukhin, Illia},
  journal={Advances in neural information processing systems},
  volume={30},
  year={2017}
}

@article{hunter2022understanding,
author = {Hunter, Elizabeth and Kelleher, John},
year = {2022},
month = {07},
pages = {100056},
title = {Understanding the assumptions of an SEIR compartmental model using agentization and a complexity hierarchy},
volume = {4},
journal = {Journal of Computational Mathematics and Data Science},
doi = {10.1016/j.jcmds.2022.100056}
}

@article{bruel2020universal,
  title={Universal function approximation on graphs},
  author={Br{\"u}el Gabrielsson, Rickard},
  journal={Advances in neural information processing systems},
  volume={33},
  pages={19762--19772},
  year={2020}
}

@article{papagiannopoulou2024long,
  title={Long-term regional influenza-like-illness forecasting using exogenous data},
  author={Papagiannopoulou, Eirini and Bossa, Mat{\'\i}as Nicol{\'a}s and Deligiannis, Nikos and Sahli, Hichem},
  journal={IEEE Journal of Biomedical and Health Informatics},
  volume={28},
  number={6},
  pages={3781--3792},
  year={2024},
  publisher={IEEE}
}

@article{lim2021temporal,
  title={Temporal fusion transformers for interpretable multi-horizon time series forecasting},
  author={Lim, Bryan and Ar{\i}k, Sercan {\"O} and Loeff, Nicolas and Pfister, Tomas},
  journal={International Journal of Forecasting},
  volume={37},
  number={4},
  pages={1748--1764},
  year={2021},
  publisher={Elsevier}
}

@inproceedings{yu2018spatio,
  title={Spatio-Temporal Graph Convolutional Networks: A Deep Learning Framework for Traffic Forecasting},
  author={Yu, Bing and Yin, Haoteng and Zhu, Zhanxing},
  booktitle={Proceedings of the Twenty-Seventh International Joint Conference on Artificial Intelligence},
  pages={3634--3640},
  year={2018},
  organization={International Joint Conferences on Artificial Intelligence Organization}
}

@inproceedings{cheng2016long,
  title={Long Short-Term Memory-Networks for Machine Reading},
  author={Cheng, Jianpeng and Dong, Li and Lapata, Mirella},
  booktitle={Proceedings of the 2016 Conference on Empirical Methods in Natural Language Processing},
  year={2016},
  organization={Association for Computational Linguistics}
}

@inproceedings{wang2022event,
  title={Event-aware multimodal mobility nowcasting},
  author={Wang, Zhaonan and Jiang, Renhe and Xue, Hao and Salim, Flora D and Song, Xuan and Shibasaki, Ryosuke},
  booktitle={Proceedings of the AAAI Conference on Artificial Intelligence},
  volume={36},
  number={4},
  pages={4228--4236},
  year={2022}
}

@inproceedings{li2018diffusion,
  title={Diffusion Convolutional Recurrent Neural Network: Data-Driven Traffic Forecasting},
  author={Li, Yaguang and Yu, Rose and Shahabi, Cyrus and Liu, Yan},
  booktitle={International Conference on Learning Representations},
  year={2018}
}

@inproceedings{deng2020cola,
  title={Cola-GNN: Cross-location attention based graph neural networks for long-term ILI prediction},
  author={Deng, Songgaojun and Wang, Shusen and Rangwala, Huzefa and Wang, Lijing and Ning, Yue},
  booktitle={Proceedings of the 29th ACM international conference on information \& knowledge management},
  pages={245--254},
  year={2020}
}

@inproceedings{xie2022epignn,
  title={EpiGNN: Exploring spatial transmission with graph neural network for regional epidemic forecasting},
  author={Xie, Feng and Zhang, Zhong and Li, Liang and Zhou, Bin and Tan, Yusong},
  booktitle={Joint European Conference on Machine Learning and Knowledge Discovery in Databases},
  pages={469--485},
  year={2022},
  organization={Springer}
}

@inproceedings{liu2023epidemiology,
  title={Epidemiology-aware deep learning for infectious disease dynamics prediction},
  author={Liu, Mutong and Liu, Yang and Liu, Jiming},
  booktitle={Proceedings of the 32nd ACM International Conference on Information and Knowledge Management},
  pages={4084--4088},
  year={2023}
}

@article{nguyen2023predicting,
  title={Predicting COVID-19 pandemic by spatio-temporal graph neural networks: A New Zealand's study},
  author={Nguyen, Viet Bach and Hy, Truong Son and Tran-Thanh, Long and Nghiem, Nhung},
  journal={arXiv preprint arXiv:2305.07731},
  year={2023}
}

@article{pu2024dynamic,
  title={Dynamic adaptive spatio--temporal graph network for COVID-19 forecasting},
  author={Pu, Xiaojun and Zhu, Jiaqi and Wu, Yunkun and Leng, Chang and Bo, Zitong and Wang, Hongan},
  journal={CAAI Transactions on Intelligence Technology},
  volume={9},
  number={3},
  pages={769--786},
  year={2024},
  publisher={Wiley Online Library}
}

@article{yu2023spatio,
  title={Spatio-temporal graph learning for epidemic prediction},
  author={Yu, Shuo and Xia, Feng and Li, Shihao and Hou, Mingliang and Sheng, Quan Z},
  journal={ACM Transactions on Intelligent Systems and Technology},
  volume={14},
  number={2},
  pages={1--25},
  year={2023},
  publisher={ACM New York, NY}
}

@inproceedings{lin2023graph,
  title={Graph Neural Network Modeling of Web Search Activity for Real-time Pandemic Forecasting},
  author={Lin, Chen and Zhou, Jianghong and Zhang, Jing and Yang, Carl and Agichtein, Eugene},
  booktitle={2023 IEEE 11th International Conference on Healthcare Informatics (ICHI)},
  pages={128--137},
  year={2023},
  organization={IEEE}
}

@inproceedings{cao2022mepognn,
  title={Mepognn: Metapopulation epidemic forecasting with graph neural networks},
  author={Cao, Qi and Jiang, Renhe and Yang, Chuang and Fan, Zipei and Song, Xuan and Shibasaki, Ryosuke},
  booktitle={Joint European conference on machine learning and knowledge discovery in databases},
  pages={453--468},
  year={2022},
  organization={Springer}
}

@inproceedings{wang2022causalgnn,
  title={Causalgnn: Causal-based graph neural networks for spatio-temporal epidemic forecasting},
  author={Wang, Lijing and Adiga, Aniruddha and Chen, Jiangzhuo and Sadilek, Adam and Venkatramanan, Srinivasan and Marathe, Madhav},
  booktitle={Proceedings of the AAAI conference on artificial intelligence},
  volume={36},
  number={11},
  pages={12191--12199},
  year={2022}
}

@inproceedings{sha2021source,
  title={Source detection on networks using spatial temporal graph convolutional networks},
  author={Sha, Hao and Al Hasan, Mohammad and Mohler, George},
  booktitle={2021 IEEE 8th International Conference on Data Science and Advanced Analytics (DSAA)},
  pages={1--11},
  year={2021},
  organization={IEEE}
}

@article{dobruschin1968description,
  title={The description of a random field by means of conditional probabilities and conditions of its regularity},
  author={Dobruschin, PL},
  journal={Theory of Probability \& Its Applications},
  volume={13},
  number={2},
  pages={197--224},
  year={1968},
  publisher={SIAM}
}

@article{jafari2021survey,
  title={A survey on locality sensitive hashing algorithms and their applications},
  author={Jafari, Omid and Maurya, Preeti and Nagarkar, Parth and Islam, Khandker Mushfiqul and Crushev, Chidambaram},
  journal={arXiv preprint arXiv:2102.08942},
  year={2021}
}

@inproceedings{datar2004locality,
  title={Locality-sensitive hashing scheme based on p-stable distributions},
  author={Datar, Mayur and Immorlica, Nicole and Indyk, Piotr and Mirrokni, Vahab S},
  booktitle={Proceedings of the twentieth annual symposium on Computational geometry},
  pages={253--262},
  year={2004}
}

@article{prosser2024using,
  title={Using an adaptive modeling framework to identify avian influenza spillover risk at the wild-domestic interface},
  author={Prosser, Diann J and Kent, Cody M and Sullivan, Jeffery D and Patyk, Kelly A and McCool, Mary-Jane and Torchetti, Mia Kim and Lantz, Kristina and Mullinax, Jennifer M},
  journal={Scientific Reports},
  volume={14},
  number={1},
  pages={14199},
  year={2024},
  publisher={Nature Publishing Group UK London}
}

@misc{GENBANK,
  author       = {National Center for Biotechnology Information (NCBI)},
  title        = {GenBank Sequence Database},
  year         = {2025},
  url          = {https://www.ncbi.nlm.nih.gov/genbank/},
  note         = {Accessed January 2025}
}

@misc{eBird_Weekly_Abundance,
  author       = {eBird},
  title        = {Weekly Bird Abundance Data},
  year         = {2022},
  url          = {https://science.ebird.org/en/status-and-trends},
  note         = {Accessed January 2025}
}

@article{kimura1980,
  title={A simple method for estimating evolutionary rates of base substitutions through comparative studies of nucleotide sequences},
  author={Kimura, Motoo},
  journal={Journal of Molecular Evolution},
  volume={16},
  number={2},
  pages={111--120},
  year={1980},
  publisher={Springer},
  doi={10.1007/BF01731581}
}

@inproceedings{zhang2019heterogeneous,
  title={Heterogeneous graph neural network},
  author={Zhang, Chuxu and Song, Dongjin and Huang, Chao and Swami, Ananthram and Chawla, Nitesh V},
  booktitle={Proceedings of the 25th ACM SIGKDD international conference on knowledge discovery \& data mining},
  pages={793--803},
  year={2019}
}

@article{hemker2024healnet,
  title={HEALNet: Multimodal fusion for heterogeneous biomedical data},
  author={Hemker, Konstantin and Simidjievski, Nikola and Jamnik, Mateja},
  journal={Advances in Neural Information Processing Systems},
  volume={37},
  pages={64479--64498},
  year={2024}
}

@inproceedings{kim2023heterogeneous,
  title={Heterogeneous graph learning for multi-modal medical data analysis},
  author={Kim, Sein and Lee, Namkyeong and Lee, Junseok and Hyun, Dongmin and Park, Chanyoung},
  booktitle={Proceedings of the AAAI Conference on Artificial Intelligence},
  volume={37},
  number={4},
  pages={5141--5150},
  year={2023}
}

@article{yu2022healthnet,
  title={HealthNet: A health progression network via heterogeneous medical information fusion},
  author={Yu, Fuqiang and Cui, Lizhen and Chen, Huanhuan and Cao, Yiming and Liu, Ning and Huang, Weiming and Xu, Yonghui and Lu, Hua},
  journal={IEEE Transactions on Neural Networks and Learning Systems},
  volume={34},
  number={10},
  pages={6940--6954},
  year={2022},
  publisher={IEEE}
}

@article{guo2023graph,
  title={Graph-based fusion of imaging, genetic and clinical data for degenerative disease diagnosis},
  author={Guo, Rui and Tian, Xu and Lin, Hanhe and McKenna, Stephen and Li, Hong-Dong and Guo, Fei and Liu, Jin},
  journal={IEEE/ACM Transactions on Computational Biology and Bioinformatics},
  volume={21},
  number={1},
  pages={57--68},
  year={2023},
  publisher={IEEE}
}

@inproceedings{hu2020heterogeneous,
  title={Heterogeneous graph transformer},
  author={Hu, Ziniu and Dong, Yuxiao and Wang, Kuansan and Sun, Yizhou},
  booktitle={Proceedings of the web conference 2020},
  pages={2704--2710},
  year={2020}
}

@article{banerjee2022spatial,
  title={Spatial--temporal synchronous graph transformer network (stsgt) for covid-19 forecasting},
  author={Banerjee, Soumyanil and Dong, Ming and Shi, Weisong},
  journal={Smart Health},
  volume={26},
  pages={100348},
  year={2022},
  publisher={Elsevier}
}

@article{kim2024forecasting,
  title={Forecasting epidemic spread with recurrent graph gate fusion transformers},
  author={Kim, Minkyoung and Kim, Jae Heon and Jang, Beakcheol},
  journal={IEEE Journal of Biomedical and Health Informatics},
  year={2024},
  publisher={IEEE}
}

@article{giacinti2024transmission,
  title={Transmission dynamics of highly pathogenic avian influenza virus at the wildlife-poultry-environmental interface: a case study},
  author={Giacinti, Jolene A and Jarvis-Cross, Madeline and Lewis, Hannah and Provencher, Jennifer F and Berhane, Yohannes and Kuchinski, Kevin and Jardine, Claire M and Signore, Anthony and Mansour, Sarah C and Sadler, Denby E and others},
  journal={One Health},
  volume={19},
  pages={100932},
  year={2024},
  publisher={Elsevier}
}

@article{pei2018forecasting,
  title={Forecasting the spatial transmission of influenza in the United States},
  author={Pei, Sen and Kandula, Sasikiran and Yang, Wan and Shaman, Jeffrey},
  journal={Proceedings of the National Academy of Sciences},
  volume={115},
  number={11},
  pages={2752--2757},
  year={2018},
  publisher={National Academy of Sciences}
}

@article{venkatramanan2021forecasting,
  title={Forecasting influenza activity using machine-learned mobility map},
  author={Venkatramanan, Srinivasan and Sadilek, Adam and Fadikar, Arindam and Barrett, Christopher L and Biggerstaff, Matthew and Chen, Jiangzhuo and Dotiwalla, Xerxes and Eastham, Paul and Gipson, Bryant and Higdon, Dave and others},
  journal={Nature communications},
  volume={12},
  number={1},
  pages={726},
  year={2021},
  publisher={Nature Publishing Group UK London}
}

@article{kandeil2023rapid,
  title={Rapid evolution of A (H5N1) influenza viruses after intercontinental spread to North America},
  author={Kandeil, Ahmed and Patton, Christopher and Jones, Jeremy C and Jeevan, Trushar and Harrington, Walter N and Trifkovic, Sanja and Seiler, Jon P and Fabrizio, Thomas and Woodard, Karlie and Turner, Jasmine C and others},
  journal={Nature communications},
  volume={14},
  number={1},
  pages={3082},
  year={2023},
  publisher={Nature Publishing Group UK London}
}

@article{caliendo2022transatlantic,
  title={Transatlantic spread of highly pathogenic avian influenza H5N1 by wild birds from Europe to North America in 2021},
  author={Caliendo, Valentina and Lewis, Nicola S and Pohlmann, Anne and Baillie, Stephen R and Banyard, Ashley C and Beer, Martin and Brown, Ian H and Fouchier, RAM and Hansen, Rowena DE and Lameris, Thomas K and others},
  journal={Scientific reports},
  volume={12},
  number={1},
  pages={11729},
  year={2022},
  publisher={Nature Publishing Group UK London}
}

\appendix

\section{Avian-US Dataset Setup}
\label{apd:dataset}
The Avian-US dataset is a spatiotemporal, multi-modal dataset designed to support forecasting and modeling of avian influenza outbreaks across the United States. It integrates epidemiological records, viral genomic sequences, and host population data across 3,227 U.S. counties from 2021–2024. Each modality is spatially and temporally aligned at the county-week level, enabling multi-layered graph construction for downstream forecasting tasks.

\subsection{Data Collection}
This dataset integrates spatiotemporal outbreak data, genomic sequences, and species-level abundance observations into a structured multilayer representation for disease forecasting. Each stream was independently collected but programmatically harmonized for modeling integration.

Infected case data were sourced from federal surveillance systems and include time-stamped infection reports for host data recorded at the county level across the continental United States from 2021 to 2024. Each record includes a free-text host descriptor, location metadata, and a collection date. To standardize taxonomic information, host descriptors were programmatically mapped to a reference taxonomy using a hierarchical classification schema derived from the International Ornithological Congress (IOC) avian taxonomy. This resolved inconsistencies such as overlapping or ambiguous common names by aligning to stable scientific identifiers.

Genomic data consist of hemagglutinin (HA) segment sequences found in publicly available viral genome repositories \cite{GENBANK}. Sequences with sufficient metadata were retained and filtered to include only wild bird hosts. A probabilistic record linkage model was used to associate sequences with case records. This model was trained on labelled match/non-match examples and used gradient-boosted decision trees to compute a match score based on taxonomic agreement, spatial proximity, and temporal overlap (within a ±14-day window). High-confidence matches were retained for downstream analysis.

Host population data were drawn from the eBird Status and Trends product, which provides weekly abundance estimates at ~3 km resolution for North American bird species \cite{eBird_Weekly_Abundance}. Raster values were extracted for each species and week, then aggregated at the county level to align with the spatial granularity of case data. Only wild bird species were retained, and abundance vectors were indexed by county and week.

All records were assigned stable identifiers and organized into structured, timestamped tables. The pipeline ensures consistency across modalities while maintaining temporal fidelity and species-level resolution.

\subsection{Data Description}
\label{apd:gene}
The dataset comprises real-world, multi-source data documenting avian influenza outbreaks in the United States from January 2021 to December 2024. It includes temporally aligned information on confirmed infection cases, viral genome sequences, and wild bird abundance estimates, collected and harmonized at a weekly resolution.

The epidemiological component consists of over 12,000 reported H5-positive wild bird cases, spanning all 48 contiguous U.S. states. Each record includes collection date, geographic location (mapped to U.S. counties), and host classification. Taxonomic labels were normalised using a hierarchical mapping system that resolves ambiguous or underspecified entries to consistent species-level identifiers, informed by IOC naming conventions.

A subset of 8,000 cases was associated with full or partial HA segment sequences retrieved from public repositories. Genomic data were filtered to retain wild bird hosts only, and sequence metadata (host, date, location) were cleaned and harmonised to match epidemiological records. Genomic divergence between HA segment sequences was computed using the K80 model, which accounts for substitution asymmetry between transitions (A$\leftrightarrow$G, C$\leftrightarrow$T) and transversions. For each aligned sequence pair, we calculate the observed proportions of transitions ($P$) and transversions ($Q$), and estimate the evolutionary distance $d$ as:
\[
d = -\tfrac{1}{2} \log(1 - 2P - Q) - \tfrac{1}{4} \log(1 - 2Q)
\]
where $P = \frac{\# \text{transitions}}{L}$ and $Q = \frac{\# \text{transversions}}{L}$, with $L$ denoting the aligned sequence length. This evolutionary distance matrix encodes biologically grounded measures of divergence under a continuous-time Markov model and is well-suited for comparing within-clade avian influenza sequences. It is used as an input feature for constructing genomic similarity edges in the downstream heterogeneous graph.

Host population context was derived from over 630 weekly avian abundance layers produced by the eBird Status and Trends project \cite{eBird_Weekly_Abundance}. These layers estimate the relative abundance of bird species at ~3 km resolution across North America. Raster values were extracted and aggregated at the county level for all species matching wild bird families in the outbreak dataset. The resulting abundance vectors were aligned weekly to match case timelines and stored in compressed array format.

All data layers were temporally aligned by epidemiological week. Metadata were standardised across data types, with fields for date, location, taxonomic label, and abundance scores. Unique identifiers were assigned to all records to enable traceability across modalities. The dataset is designed to support temporal, ecological, and genetic analysis of avian influenza dynamics in wild bird populations using real-world observations, without reliance on synthetic or simulated data.

\section{Experiments}
\subsection{Experimental Settings.}
\label{apd:setting}
In our empirical evaluations, we implement ST-GCN, SelfAttnRNN, DCRNN, and Cola-GNN using the open-source Cola-GNN repository~\footnote{https://github.com/amy-deng/colagnn}.
ST-Net and EAST-Net are built upon the official EAST-Net implementation~\footnote{https://github.com/underdoc-wang/EAST-Net/tree/main}. 
Implementation of Epi-GNN~\footnote{https://github.com/Xiefeng69/EpiGNN} and Epi-Cola-GNN~\footnote{https://github.com/gigg1/CIKM2023EpiDL/tree/main} are based on their respective publicly available source code.

Except HGT, all bas are primarily designed for single-layer spatio-temporal forecasting and assume either a fixed or learnable graph structure. 
As such, they are not directly compatible with the multi-layer architecture of the Avian-US dataset. 
To make them applicable, we adapt each model by building homogeneous graphs tailored to its design.
For ST-GCN, SelfAttnRNN, Cola-GNN, EpiGNN, EpiCola-GNN, and DCRNN, we define a binary adjacency matrix based on spatial proximity and genetic similarity between locations. 
This setup mirrors their original use cases, allowing these models to learn spatio-temporal patterns over a fixed location-level graph.
For ST-Net and EAST-Net, which support adaptive graph learning, we initialize homogeneous graphs with no predefined edges. 
These models learn the graph structure dynamically, allowing them to infer inter-location dependencies during training without relying on geographic priors.

We set a short observation window of $T$=4 steps and forecast the next $H$=4 steps (all experiments are conducted under this setting unless specified), and report the averaged evaluation metrics of $H$ steps. 
Experiments of all baselines and \ourmethod~are conducted under a 5-fold cross-validation with the same random seed to ensure consistency.
In addition to fixed embedding size $d$=8 and weight regularization $\lambda_2=5e-4$, all baseline models are re-trained and tuned for optimal performance using their official open-source code. 
For \ourmethod~, we search $\lambda_1 \in \{0.01, 0.05, 0.1, 0.5, 1\}$, and choose $\lambda_1=0.1$ for final evaluations.
All experiments are run on either a single NVIDIA V100, DGX A100, or NVIDIA RTX A5000 GPU.

To ensure comparability in overall performance, we unify the embedding size and hidden dimensions across all models.
We further apply the stratified weighting infection loss in Section \ref{sec:opt} to all baselines and re-run each baseline on the Avian-US dataset for fair comparison. 
The key hyperparameters used for all baseline models are listed below:

\noindent{\textbf{ST-GCN}}:
embedding size=8, hidden dim=16, number of layers=3, epoch=100, learning rate=1e-5, dropout=0.3, window size=4, predicted horizon=4, weight of regularization term=5e-4.

\noindent{\textbf{SelfAttnRNN}}:
embedding size=8, hidden dim=16, number of layers=2, epoch=100, learning rate=1e-5, dropout=0.3, window size=4, predicted horizon=4, weight of regularization term=5e-4.

\noindent{\textbf{DCRNN}}:
embedding size=8, hidden dim=16, number of layers=2, max step of random walk=3, epoch=100, learning rate=1e-5, dropout=0.3, window size=4, predicted horizon=4, weight of regularization term=5e-4.

\noindent{\textbf{Cola-GNN}}:
embedding size=8, hidden dim=16, number of filter=10, dilated rate for short term=1, dilated rate for long term=2, epoch=100, learning rate=1e-5, dropout=0.3, number of RNN layers=1, number of GNN layers=2, window size=4, predicted horizon=4, weight of regularization term=5e-4.

\noindent{\textbf{ST-Net}}:
embedding size=8 (data) and 8 (time), Chebyshev layers=3, encoder layer=2, decoder layer=2, epoch=100, learning rate=1e-5, dropout=0.3, window size=4, predicted horizon=4, weight of regularization term=5e-4.

\noindent{\textbf{EAST-Net}}:
spatial embedding size=8, modality embedding size =4, time embedding size=8, mobility prototype number = 8, memory dimension = 16, Chebyshev layers=3, encoder layer=2, decoder layer=2, epoch=100, learning rate=1e-5, dropout=0.3, window size=4, predicted horizon=4, weight of regularization term=5e-4.

\noindent{\textbf{Epi-GNN}}:
embedding size=8, hidden dim=16, hidden dim of attention layer=64, pooling layer=2, patience=100, GNN layers=2, filer size=$\mathbf{f}_{1 \times 5,1}$ and $\mathbf{f}_{1 \times 3,1}$, window size=4, predicted horizon=4, epoch=100, learning rate=1e-5, dropout=0.3, weight of regularization term=5e-4.

\noindent{\textbf{Epi-Cola-GNN}}:
embedding size=8, hidden dim=16, weight of epidemiological loss=0.5, patience=150, epoch=100, learning rate=1e-5, dropout=0.3, window size=4, predicted horizon=4, weight of regularization term=5e-4.

\noindent{\textbf{STSGT}}:
embedding size=8, hidden dim=16, number of STSGT layers$=2$, number of head$=2$, dropout rate$=0.3$, sampling number $n=128$, sampling depth $L=2$, learning rate=0.001, window size=4, predicted horizon=4, weight of regularization term=1e-4.

\noindent{\textbf{HGT}}:
embedding size=8, hidden dim=16, number of layers$=2$, dropout rate$=0.3$, sampling number $n=128$, sampling depth $L=2$, learning rate=0.001, window size=4, predicted horizon=4, weight of regularization term=1e-4.

\noindent{\textbf{BLUE}}:
embedding size=8, construction smoothing layer $K$=2, $B$=10, GraphSAGE layers $L$=2, $\lambda_1$=0.9, $\lambda_2$=5e-4, epoch=100, learning rate=1e-5, dropout=0.3, window size=4, predicted horizon=4, $\lambda_o$=0.3, $\lambda_p$=0.3.
The infection severities are set as $\tau_{low}=1$, $\tau_{med}=5$, $\tau_{high}=25$, the corresponding weight are $w_{low}=1.0$, $w_{med}=8.0$, $w_{high}=15.0$.

\subsection{Impact of $\lambda_o$}
In the decoder process, $\lambda_o$ controls the influence of the previously decoded state.
To isolate the impact of this temporal recurrence, we fix the observation and prediction windows at $T=4$ and $H=4$, respectively, and maintain the encoder hidden state weight at $\lambda_p=0.3$.
We then evaluate the model's performance by varying $\lambda_o$ within the set $\{0.1, 0.2, 0.3, 0.4, 0.5, 0.6\}$.
The results are summarized in Table.\ref{apd:tab-prev}.

Lower values of $\lambda_o$ prioritize the immediate features of the current step, whereas higher values enforce stronger continuity with previous decoder outputs. Empirically, we observe that MAE is minimized at $\lambda_o=0.3$, while RMSE achieves its lowest at $\lambda_o=0.5$.
Ranking metrics (F1, PCC, SCC) demonstrate a unimodal distribution: they gradually improve as $\lambda_o$ increases, peaking around the mid-range, before significantly deteriorating. This suggests that while a moderate reliance on previous predictions aids in capturing long-horizon trends, excessive autoregressive weighting ($\lambda_o > 0.4$) dilutes the informative signals from the current step, leading to error propagation. Consequently, we select $\lambda_o=0.3$ as the optimal equilibrium, balancing the need for robust long-range trend extraction with the precision required for accurate infection count prediction.

\begin{table}[t]
\centering
\caption{impact of $\lambda_o$ on the Avian-US dataset.}
\label{apd:tab-prev}
\begin{tabular}{l|ccccc}
\toprule
$\lambda_o$ & RMSE   & MAE    & F1     & PCC    & SCC    \\ \hline
0.1         & 0.6382 & 0.0921 & 0.0945 & 0.0717 & 0.1157 \\
0.2         & 0.6274 & 0.0889 & 0.0861 & 0.0762 & 0.1121 \\
0.3         & {\ul 0.6106} & \textbf{0.0848} & \textbf{0.1001} & \textbf{0.0871} & \textbf{0.1217} \\
0.4         & 0.6112 & 0.0877 & 0.0906 & {\ul 0.0854} & {\ul 0.1201} \\
0.5         & \textbf{0.6006} & {\ul 0.0854} & {\ul 0.0924} & 0.0844 & 0.1183 \\
0.6         & 0.6132 & 0.0897 & 0.0825 & 0.0853 & 0.1194 \\ \bottomrule
\end{tabular}
\end{table}

\begin{table}
\centering
\caption{Different predicted horizon $H$ under window size $T=4$ (up) and $T=8$ (down).}
\vspace{-2mm}
\label{apd:tab-avian}
    \resizebox{\linewidth}{!}{
    \begin{tabular}{l|cccc}
    \toprule
    $T$     & \multicolumn{4}{c}{4}                             \\ \hline
    $H$     & 1               & 2               & 4               & 8             \\ \hline
    RMSE    & 0.6018 ± 0.0730 & 0.6098 ± 0.0653 & 0.6106 ± 0.0719 & 0.6174 ± 0.0319\\
    MAE     & 0.0955 ± 0.0217 & 0.0924 ± 0.0198 & 0.0848 ± 0.0182 & 0.0961 ± 0.0358\\
    F1      & 0.1271 ± 0.0166 & 0.1109 ± 0.0273 & 0.1001 ± 0.0193 & 0.0949 ± 0.0113\\
    PCC     & 0.1195 ± 0.0339 & 0.0976 ± 0.0188 & 0.0871 ± 0.0132 & 0.0683 ± 0.0074\\ 
    SCC     & 0.1489 ± 0.0148 & 0.1345 ± 0.0230 & 0.1217 ± 0.0140 & 0.1029 ± 0.0066\\ \bottomrule
    \end{tabular}
    }

    \resizebox{\linewidth}{!}{
    \begin{tabular}{l|cccc}
    \toprule
    $T$     & \multicolumn{4}{c}{8}                                                 \\ \hline
    $H$     & 1               & 2               & 4               & 8               \\ \hline
    RMSE    & 0.6337 ± 0.0394 & 0.6112 ± 0.0411 & 0.6268 ± 0.0292 & 0.6178 ± 0.0434 \\
    MAE     & 0.1151 ± 0.0200 & 0.0952 ± 0.0245 & 0.1061 ± 0.0383 & 0.1065 ± 0.0212 \\
    F1      & 0.1226 ± 0.0169 & 0.1269 ± 0.0175 & 0.1030 ± 0.0074 & 0.0950 ± 0.0069 \\
    PCC     & 0.1225 ± 0.0237 & 0.1084 ± 0.0151 & 0.0828 ± 0.0208 & 0.0788 ± 0.0088 \\ 
    SCC     & 0.1587 ± 0.0158 & 0.1472 ± 0.0085 & 0.1208 ± 0.0161 & 0.1172 ± 0.0127 \\ \bottomrule
    \end{tabular}
    }
\vspace{-3mm}
\end{table}

\subsection{Impacts of \texorpdfstring{$T$}{T} and \texorpdfstring{$H$}{H}}
\label{sec:parameter}
We evaluate the robustness of \ourmethod~by varying the observation window size $T \in \{4,8\}$ and the prediction horizon $H \in \{1,2,4,8\}$.
\paragraph{Impact of Prediction Horizon $H$.}
\ourmethod~performs best at $H=1$, indicating that short-term temporal dependencies in the Avian-US dataset are significantly more stationary and predictable than long-term dynamics. 
As $H$ increases, ranking metrics (F1, PCC, SCC) decline due to the unpredictability of long-term transmission trends.
The irregularity in avian influenza outbreaks, potentially due to \textbf{variations in viral infectivity} across different strains, makes long-term forecasting challenging.
The significant performance drop at $H=8$ validates the \textbf{error accumulation} problem inherent to the autoregressive decoding process, indicating a loss of predictive accuracy when the forecast horizon is too long.

\paragraph{Impact of Observation Window $T$.}
Considering a fixed prediction horizon $H$, increasing the observation window from $T=4$ to $T=8$ shows a consistent positive trend:
(i) \textbf{Enhanced Pattern Capture}: Longer observation windows result in a consistent decrease in RMSE and MAE, along with an upward trend in F1/PCC/SCC. This indicates that access to extended observations allows \ourmethod to better capture underlying patterns that short windows miss.
(ii) \textbf{Increased Variability}: the performance variances of nearly all metrics increase with larger $T$. 
This is likely because longer observation windows, especially with sparse data, introduce a larger proportion of zero-valued intervals, which enlarges variability across validation folds and leads to greater fold-to-fold variability in results.

\end{document}